\documentclass[amsmath,superscriptaddress,amssymb,aps,prd,floatfix,twocolumn,nofootinbib]{revtex4-2}

\usepackage[utf8]{inputenc}
\usepackage{epsfig}
\usepackage{graphicx}
\usepackage{mathrsfs}
\usepackage[bookmarks=false,hidelinks]{hyperref}
\usepackage{dcolumn}
\usepackage{bm}
\usepackage{float}
\usepackage{soul}
\usepackage{amsmath}
\usepackage{bbold}

\usepackage{placeins}

\allowdisplaybreaks

\newcommand{\ud}{\mathrm{d}}
\newtheorem{theorem}{Theorem}

\newcommand{\tlt}[2]{\begin{tabular}[c]{@{}c@{}}#1\\ #2\end{tabular}}

\begin{document}

\title{Revisiting the quasinormal modes of the Schwarzschild black hole: Numerical analysis}

\author{Luis A. H. Mamani}%
\email{luis.mamani@uemasul.edu.br}
\affiliation {Centro de Ci\^encias Exatas Naturais e Tecnológicas,\\ Universidade Estadual da Regi\~ao Tocantina do Maranh\~ao,\\ Rua Godofredo Viana 1300, 65901- 480, Imperatriz, Maranhão, Brazil}

\author{Angel D. D. Masa}%
\email{angel.masa@ufabc.edu.b}
\affiliation {Centro de Ci\^encias Naturais e Humanas, Universidade
Federal do ABC, Avenida dos Estados 5001, 09210-580 Santo Andr\'e, São Paulo,
Brazil}

\author{Lucas Timotheo Sanches}
\email{lucas.t@ufabc.edu.br}
\affiliation {Centro de Ci\^encias Naturais e Humanas, Universidade
Federal do ABC, Avenida dos Estados 5001, 09210-580 Santo Andr\'e, São Paulo,
Brazil}

\author{Vilson T. Zanchin}%
\email{zanchin@ufabc.edu.br}
\affiliation {Centro de Ci\^encias Naturais e Humanas, Universidade
Federal do ABC, Avenida dos Estados 5001, 09210-580 Santo Andr\'e, São Paulo,
Brazil}

\begin{abstract}
We revisit the problem of calculating the quasinormal modes of spin $0$, $1/2$, $1$, $3/2$, $2$, and spin $5/2$ fields in the asymptotically flat Schwarzschild black hole spacetime. Our aim is to investigate the problem from the numerical point of view, by comparing some numerical methods available in the literature and still not applied for solving the eigenvalue problems arising from the perturbation equations in the Schwarzschild black hole spacetime. We focus on the pseudo-spectral and the asymptotic iteration methods. These numerical methods are tested against the available results in the literature, and confronting the precision between each other.  
Besides testing the different numerical methods, we calculate higher overtones quasinormal frequencies for all the investigated perturbation fields in comparison with the known results. In particular, we obtain purely imaginary frequencies for spin $1/2$ and $3/2$ fields that are in agreement with analytic results reported previously in the literature. The purely imaginary frequencies for the spin $1/2$ field are exactly the same as the frequencies obtained for the spin $3/2$ field. In turn, the quasinormal frequencies for the spin $5/2$ perturbation field are calculated for the very first time, and purely imaginary frequencies are found also in this case. We conclude that both methods provide accurate results and they complement each other.
\end{abstract}

\maketitle

\section{Introduction}
\label{introd}

Perturbation theory is a  very useful theoretical toolkit for the investigation of properties of a physical system. For example, the stability under small perturbations. In the harmonic oscillator problem, it drives into a second-order differential equation with Dirichlet boundary conditions whose solutions are characterized by a set of discrete real frequencies, i.e., normal modes (NMs). However, there are physical systems whose boundary conditions drive solutions with complex frequencies, i.e., quasinormal modes (QNMs), for example, a harmonic oscillator into a dissipative medium, see for instance \cite{goldstein:mechanics}. Thus, the investigation of the quasinormal (QN) frequencies and its mathematical properties become fascinating subject which may shed light into the understanding of universal properties of the physical system under investigation.

In the context of gravity theories, perturbation theory is important for several reasons.
One of the motivations is the investigation of gravitational waves spectroscopy \cite{LIGOScientific:2017vwq}. One also may use perturbation theory to investigate the stability under small perturbations of determined background solutions. It has been shown that the perturbation equations may be written as a second-order differential equations allowing us to use numerical techniques implemented in differential equations to solve them. One of the boundary conditions considers that classically noting comes out from the black hole interior, so that the boundary condition at the horizon are ingoing waves. In turn, at the spatial infinite the condition are outgoing waves, because nothing can come from outside of the spacetime. Solving the perturbation equations under these particular boundary conditions drive to solutions with discrete (complex) eigenvalues. Such eigenvalues are frequencies representing the characteristic oscillations of the black hole that relaxes 
after being perturbed. One very interesting property of these frequencies is that they do not depend on the initial perturbation and are fixed completely by the properties of the black hole under consideration. 

The quasinormal modes are of particular interest in black hole astrophysics.  Direct observations showed in the coalescence of a binary system emits gravitational waves in the form of QNMs, that is, the final system obeys the predictions of black hole perturbation theory~\cite{Abbott:2016}. This fact alone is enough to motivate the development, testing, and comparison of different methods for finding QNMs, but teh relevance o QNMs go far further this fact. For reviews and additional discussions about QNMs in different contexts see, for instance, \cite{Berti:2009, Konoplya:2011qq, York:1983zb, Ferrari:1984zz, Leaver:1986gd, Nollert:1992ifk, Andersson:1992scr, Nollert:1993zz, Kokkotas:1999bd, Horowitz:1999jd, Nollert:1999ji, Cardoso:2001hn, Cardoso:2001bb, Konoplya:2002zu, Starinets:2002br, Clarkson:2002jz, Kovtun:2005ev, Cardoso:2008bp, Miranda:2008vb, Morgan:2009pn, Miranda:2009uw, Mamani:2013ssa, Mamani:2018qzl,Mamani:2018uxf, Mamani:2022qnf} and references therein.

In turn, the first observation of the shadow of the supermassive black hole M87* by the Event Horizon Telescope Collaboration (EHT) \cite{EventHorizonTelescope:2019dse, EventHorizonTelescope:2019ggy} open a new window for the investigation of strong gravitational field phenomena. 
There is also the studies proposing a connection between the real part of the quasinormal (QN) frequency with the radius of the shadow, see the Refs.~\cite{Perlick:2015vta, Bisnovatyi-Kogan:2017kii, Cuadros-Melgar:2020kqn, Jusufi:2020mmy, Jusufi:2020dhz}. Among the new possibilities, there are researched papers attempting to find constraints in parameters arising in different models like in general uncertainty principle (GUP), see for instance  Ref.~\cite{Neves:2020doc, Neves:2019lio}.

In this paper we are going to revisit the calculation of quasinormal modes for integer spin $0$, $1$ and $2$ fields, as well as for semi-integer spin $1/2$, $3/2$ and $5/2$ fields. Since Chandrasekhar calculated the quasinormal modes for $s=2$ in Ref.~\cite{chandrasekhar1975quasi}, the problem of calculating QNMs for other spin fields were previously investigated in the literature using different techniques (numeric and analytic), see for instance Refs.~\cite{leaver1985analytic, Cho:2003qe, Shu:2005fw, Konoplya:2004ip, Cho:2011sf}. However, our approach here is from the numerical point of view, for doing so we are going to use two numerical methods well established in the literature. The first one is the pseudo-spectral method used to solve differential equations expanding the solution in a base composed by special functions \cite{boyd01}. This method was used to calculate the quasinomal modes of Schwarzschild black hole for spin zero field in Ref.~\cite{Jansen:2017oag}. However, we extend the method for calculating the QNMs for spin $1/2$, $1$, $3/2$, $2$ and $5/2$ fields. We also use the asymptotic iteration method (AIM) proposed originally in Ref.~\cite{Ciftci:2003}. This method was extended to solve quasinormal modes in Ref.~\cite{Cho:2011sf}. In this paper we 
review the relatively unexplored asymptotic iteration method and apply it to the QNM problem. We also introduce a new software package that implements the latter method for usage in general second order ODEs.

This paper is organized as follows. In Section \ref{Sec:Equations} we write the equations of motion describing the spin $0$, $1/2$, $1$, $3/2$, $2$ and $5/2$ fields in a suitable form to apply the numerical methods. In Section \ref{Sec:PseudoSpectral} we review and discuss the pseudo-spectral method, we focus in the way how this method can be applied, by expanding the solution using one or two special functions. Section \ref{Sec:AIM} is devoted to discuss about the AIM and it extension to calculate QN frequencies. We also present a open source code that can be used freely. In turn, in Section \ref{Sec:NumericalResults} we present our numerical results obtained by both methods, we also compare against numerical results available in the literature. We leave the discussion of the QNMs in the limit of large angular for Section \ref{Sec:QNMsLargel}, were we also compare against analytic results. Finally, our main conclusions are presented in Section \ref{Sec:Conclusions}. Additional details are presented in Appendices \ref{Sec:Spin5f2Field} and \ref{Sec:AddDiscu}.

\section{Equations of motion}
\label{Sec:Equations}

In this section we write the equations describing the field perturbations on the gravitational background solution of the Einstein equations. We focus on the metric for an spherical symmetric black hole, which is given by \cite{Schwarzschild:1916uq}
\begin{equation}\label{EqMetric}
ds^2=-f(r)\,dt^2+\frac{1}{f(r)}dr^2+r^2d\theta^2+
r^2\sin^2{\theta}\,d\varphi^2,
\end{equation}
where the horizon function of the Schwarzschild black hole is given by 
\begin{equation}\label{EqHoriFuncHayward}
f(r)=1-\frac{2M}{r}.
\end{equation}
where $M$ is the mass of the black hole, and $r$ is the radial coordinate which, in principle, belongs to the interval $r\in [0,\, \infty)$. The coordinates in the metric \eqref{EqMetric} are known as Schwarzschild coordinates. As it is well known, this metric presents an event horizon at $r=2M$ and a curvature (physical) singularity at $r=0$. In the asymptotic region, i.e., $r\to \infty$, the metric reduces to flat metric. As long as the quasinormal modes in this black hole spacetime are concerned, the interesting region is the spacetime region spanned by the radial coordinate $r$ in the interval $2M< r<\infty$.

\subsection{Spin $0$, $1$ and $2$ perturbations}

Here we revisit the study of perturbations of integer spin, such that scalar, vector, and gravitational perturbations in the Schwarzschild black hole peacetime.  This is a long standing problem, and there are a considerable amount of results published in the literature, what is certainly interesting for the purpose of the present work.
In fact, it was proven that the equations of motion can be written in a compact form, the so called, Schr\"odinger-like differential equations, see for instance \cite{Berti:2009}. Thus, for massless scalar ($s=0$), electromagnetic ($s=1$) and vector type gravitational perturbations ($s=2$), the Schr\"odinger-like equations are given by
\begin{equation}\label{Eq:IntegerSpin}
\frac{d^2 \psi_{{s}}(r)}{d r_*^2}
+\left[\omega^2-V_{{s}}(r)\right]\psi_{{s}}(r)=0.
\end{equation}
where the potential is given by
\begin{equation}\label{Eq:IntegerSpinPot}
V_{s}(r)=f(r)\left(\frac{\ell(\ell+1)}{r^2}
+(1-s^2)\frac{2M}{r^3}\right),
\end{equation}
where the tortoise coordinate is defined in terms of the areal coordinate $r$ by $dr_*=dr/f(r)$. So far, the problem of calculating quasinormal modes was reduced to solve an eigenvalue problem. We will see that it is possible to solve this problem following two approaches, one of them expanding the function $\psi$ in a base composed by especial functions, while the other solving directly the second-order differential equation.

On the other hand, note that the potential \eqref{Eq:IntegerSpinPot} is zero at the horizon, $f(r_h)=0$. Thus, the Schr\"odinger-like equation reduces to a single harmonic oscillator problem, whose solutions are:
\begin{equation} \label{eq:sol1}
\psi_{s}(r)=c_1\, e^{-i\omega r_*}+c_2\, e^{i\omega r_*}, \quad r\to r_h.
\end{equation}
The first of these solutions is interpreted as an ingoing wave, i.e., a wave that travels inward and eventually falls into the black hole event horizon. The second solution is interpreted an outgoing wave, i.e., a wave that travels outward with respect to the black hole and can escape to space infinity. Waves travelling as this second solution would represent waves coming  from the interior of the black hole. Since the perturbation theory is implemented using classical assumptions, nothing is expected to come out from the black hole interior, thus, in the following analysis we impose the first solution as boundary condition at the horizon, we set $c_2=0$.

We also need to investigate the spatial infinity, at $r\to\infty$, where $f(r)\to 1$ and the effective potential \eqref{Eq:IntegerSpinPot} also vanishes. Thus, in such a limit the general solutions to the wave equation \eqref{Eq:IntegerSpin} have the same form as the function given in Eq.~\ref{eq:sol1}, i.e.
\begin{equation}
\psi_{s}(r_*)=c_3\, e^{-i\omega r_*}+c_4\, e^{i\omega r_*}, \quad r\to \infty.
\end{equation}
The first solution is interpreted physically as waves coming in from outside the universe and must be avoided setting $c_3=0$. In turn, the second solution represent waves going out the universe, this is the boundary condition at the spatial infinite. Finally, note that the boundary conditions do not depend explicitly on the angular momentum $\ell$ neither the spin.

It is interesting to see the behavior of the tortoise coordinate, which close to the horizon is given by ($r_h=2M$)
\begin{equation}
r_*=\int \frac{dr}{f'(r_h)(r-r_h)}\approx \frac{\ln{(r-r_h)}}{ f'(r_h)},\quad r\to r_h
\end{equation}
Thus, in terms of the radial coordinate the boundary condition at the horizon becomes ($f'(r_h)=1/r_h$)
\begin{equation}
\psi_s(r)\sim e^{-i\omega \frac{\ln{(r-r_h)}}{f'(r_h)}}\sim \left(r-r_h\right)^{-i\frac{\omega}{f'(r_h)}}.
\end{equation}
In turn, the tortoise coordinate at the spatial infinite becomes
\begin{equation}
r_*=\int \frac{dr}{f(r)}\approx r+r_h\,\ln{r},\quad r\to \infty
\end{equation}
while the asymptotic solution at the spatial infinite becomes
\begin{equation}
\psi_s(r)\sim e^{i\omega(r+ r_h \ln{r})}\sim r^{i\, r_h\omega}e^{i\omega r}.
\end{equation}

In the following, we are going to change our strategy and use a new coordinate defined by $u=2M/r$. This is equivalent to the choice $u=1/r$ and then normalizing the mass $M$ to $2M=1$. 
The relation between the tortoise coordinate $r_*$ and the new coordinate $u$  becomes $du/dr_*=-u^2f(u)$.

We also constraint our analysis to the outer region of the black hole, such that $r_h\leq r<\infty$. Hence, in terms of the new coordinate this region is bounded to the interval $u\in [0,1]$, and the potential becomes
\begin{equation}
V_{s}(u)=f(u)\,u^2\left[\ell\left(1+\ell\right)+\left(1-s^2\right)\,u\right], 
\end{equation}
where $f(u) = 1- \,u$.

To implement the pseudo-spectral method, one must write the equations on the background metric written terms of the Eddington-Filkenstein coordinates, see for instance the discussion in Ref.~\cite{Jansen:2017oag}. However, we found a short way to write the equations going directly from the Schrodinger-like equation by implementing some transformations. At the end, this transformations lead to a differential equation which is the same as the one obtained from the metric in Eddington-Filkenstein coordinates.
In order to write the perturbation equations in terms of the Eddington-Filkenstein coordinate we must implement the transformation $\psi_{s}\to \Phi_s$ given by
\begin{equation}\label{Eq:TransN1}
\psi_{s}= \frac{\Phi_{s}(u)}{u}e^{-i\omega\,r_*(u)}.
\end{equation}

Thus, Eq.~\eqref{Eq:IntegerSpin} becomes
%
\begin{equation}\label{Eq:IntegerSpin2}
\begin{split}
&\left[s^2u^2-\ell(\ell+1)u-2\,i\,\omega\right]\Phi_{s}(u)\\
&-u\left(u^2-2\,i\,\omega\right)\Phi_{s}'(u)+\left(1-u\right)\,u^3\,\Phi_{s}''(u)=0,\end{split}
\end{equation}
%
where we have set $M=1/2$ so that $r_h=1$. The asymptotic solutions close to the horizon may be calculated using the ansatz $\Phi_{s}(u)=(1-u)^{\alpha}$. By substituting this ansatz into \eqref{Eq:IntegerSpin2} we get two solutions,
\begin{equation}\label{Eq:AsympHorion}
\alpha=0,\qquad\qquad \alpha=2\,i\,\omega.
\end{equation}
The solution for $\alpha =0$ is interpreted physically as the ingoing waves at the horizon, while the other one is interpreted as a wave coming out from the black hole interior. Therefore, the second solution is neglected in the following analysis.

In the same way, we consider the ansatz $\Phi_{s}=u^{\beta}$ to get the asymptotic solution close to the spatial infinite. Plugging this ansatz in \eqref{Eq:IntegerSpin2} we get the following solution,
\begin{equation}\label{Eq:AsymInfinite}
\Phi_{s}(u)=c_5\,e^{2\,i\,\omega/u}u^{-2\,i\,\omega}+c_6\,u.
\end{equation}
We want the divergent solution, such that we set $c_6=0$. Then, we implement the final transformation which takes into consideration the boundary conditions,
\begin{equation}\label{Eq:FinalTrans}
\Phi_{s}(u)=e^{2\,i\,\omega/u}u^{-2\,i\,\omega}\phi_{s}(u),
\end{equation}
where $\phi_{s}(u)$ is a regular function in the interval $u\in[0,\,1]$ by definition. Finally, the equation of motion describing spin $0$, $1$ and $2$ perturbations is given by
%
\begin{equation}\label{Eq:IntegerSpin3}
\begin{split}
&\Big[\ell(\ell+1)\, u -s^2u^2-4\,i\lambda-16\,u(1+u)\lambda^2\Big]\phi_{s}(u)+\\
& \Big[u^3+4\,i\,u\left(1-2u^2\right)\lambda\Big]\phi_{s}'(u)-(1-u)u^3\phi_{s}''(u)=0,
\end{split}
\end{equation}
%
where we have used $\lambda=\omega M=\omega/2$. The final differential equation is then a quadratic eigenvalue problem in $\lambda$. It is worth also mentioning that in the limit of zero spin $s\to 0$, Eq.~\eqref{Eq:IntegerSpin3} reduces to Eq.~(4.8) of Ref.~\cite{Jansen:2017oag}. These results just proves that the alternative way for getting the equations for the integer spin perturbations presented here is consistent with other approaches found in the literature.

The differential equation \eqref{Eq:IntegerSpin3} was solved numerically by means of the pseudo-spectral and AIM methods. The results for perturbations of spin $0$, $1$, and $2$ are presented in  Secs.~\ref{sec:numspin0}, \ref{sec:numspin2}, and ~\ref{sec:numspin2}, respectively, where a comparison with the corresponding data in the literature is also performed. 

In the following we extend the analysis of the present section to other kind of perturbations.

\subsection{Spin $1/2$ perturbations}
\label{Sec:Spin1f2}

For half-integer spin perturbations the history is different, the differential equations are quite distinct from \eqref{Eq:IntegerSpin3}.  The equation for the spin 1/2 Dirac field as a perturbation on the Schwarzschild background was derived in Ref.~\cite{Cho:2003qe} by using the Newman-Penrose formalism. The analysis was generalized for arbitrary half-integer spin in Ref.~\cite{Shu:2005fw}. The resulting equation of motion for the perturbations may be written in the Schr\"odinger-like form Eq.~\eqref{Eq:IntegerSpin}, where the potential for the massless spin 1/2 field is given by
\begin{equation}\label{eq:pot-s12}
V_{\scriptscriptstyle{1/2}}= \frac{\left(1+\ell\right)\sqrt{f(r)}}{r^{2}} \left[\left(1+\ell\right)\sqrt{f(r)}+\frac{3M}{r}-1\right].
\end{equation} 
It is worth mentioning that we have found a typo in the definition of $\Delta$ in \cite{Cho:2003qe}, which must be $\Delta=r(r-2M)$.

We then implement the same transformations done in the integer spin cases. First, we change the radial coordinate to $u=2M/r$, which is defined in the interval $u\in[0,1]$. Then by setting $2M=1$ the effective potential \eqref{eq:pot-s12} becomes
\begin{equation}
V_{\scriptscriptstyle{1/2}}=(1+\ell)u^2\sqrt{f(u)} \left[\left(1+\ell\right)\sqrt{f(u)}+\frac32 u -1\right].
\end{equation}

It is interesting pointing out that the asymptotic solutions do not depend on the spin of the field, for that reason the asymptotic solutions for this problem are the same as those obtained in Eqs.~\eqref{Eq:AsympHorion} and \eqref{Eq:AsymInfinite}. Then, similar transformations as those giving in Eqs.~\eqref{Eq:TransN1} and~\eqref{Eq:FinalTrans} can be applied also in the present spin 1/2 case. Thus, the differential equation to be solved is given by 
\begin{equation}\label{eq:onda-s12}
\begin{split}
&R(u)\phi_{\scriptscriptstyle{1/2}}(u) + Q(u)\phi_{\scriptscriptstyle{1/2}}'(u) + P(u) \,\phi_{\scriptscriptstyle{1/2}}''(u)=0,
\end{split}
\end{equation}
in which the coefficients $R(u)$, $Q(u)$, and $P(u)$ are given by
\begin{eqnarray}
R(u)=\,&& u^3+u(1+\ell)\left(1+\ell-\sqrt{1-u}\, \right) \nonumber \\ 
&& +\frac{u^2}{2}\left[(1+\ell)\left(3\sqrt{1-u}-4\right)-2\ell^2\right]\nonumber \\
&&-4\,i\,(1-u)\lambda-16u(1-u^2)\lambda^2,\\
Q(u) = && u^3(1-u)+4\,i\,u\, \lambda\left(1-u-2u^2+2u^3\right), \\
 P(u)  =&& -u^3(1-u)^2,
 \end{eqnarray}
respectively, and with $\lambda$ standing for $\lambda=M\omega=\omega/2$.

As it can be seen, the differential equation \eqref{eq:onda-s12} is impregnated by square roots that may difficult the convergence of the numerical methods. To avoid the square roots we implement an additional change of variable $\chi^2=1-u$. Nevertheless, the new coordinate also belongs to the interval $\chi \in [0,\,1]$. The differential equation \eqref{eq:onda-s12} becomes
\begin{equation}\label{eq:onda-s12a}
\begin{split}
&R(\chi)\phi_{\scriptscriptstyle{1/2}}(\chi) + Q(\chi)\phi_{\scriptscriptstyle{1/2}}'(\chi) + P(\chi) \,\phi_{\scriptscriptstyle{1/2}}''(\chi)=0,
\end{split}
\end{equation}
in which the coefficients $R(\chi)$, $Q(\chi)$, and $P(\chi)$ are given by
\begin{eqnarray} 
R(\chi)=
\,&&2(1-\chi^2)\left[\left(\ell+1\right) \left(1+ 2\ell\,\chi-3\chi^2\right) + 2\ell\, \chi +2\chi^3\right] \nonumber \\ 
&& - 8\,i\,\chi \,\lambda-32\,\chi\left(2-3\chi^2+\chi^4\right)\lambda^2,\\
Q(\chi) = && (\chi^2-1\big)\left[\big(1-\chi^2\big)^2-8\,i\big(1-4\chi^2+2\chi^4\big)\lambda\right],\\
 P(\chi)  =&& -\chi\big(1-\chi^2\big)^3.
 \end{eqnarray}
In Sec.~\ref{sec:numspin1/2} we solve Eq.~\eqref{eq:onda-s12a} by
using the pseudo-spectral and AIM methods and compare our results against the results of Refs.~\cite{Cho:2003qe, Shu:2005fw}.

\subsection{Spin $3/2$ perturbations}

As well as for spin $1/2$ perturbation, the perturbation equation for spin $3/2$ field is different. To get the equation we are going to use the result obtained in Ref.~\cite{Shu:2005fw}, specifically Eq.~(37) of this reference, setting $s=3/2$ on this equation we get the potential of the Schr\"odinger-like equation \eqref{Eq:IntegerSpin}
\begin{equation} \begin{split}
V_{\scriptscriptstyle{3/2}} &=\frac{(1+\ell)(2+\ell)(3+\ell)\sqrt{f(r)}}{\left[2M+r(1+\ell)(3+\ell)\right]^2} \bigg(\frac{2M^2}{r^2}\\ &+(1+\ell)(3+\ell)\left[(2+\ell)\sqrt{f(r)}+\frac{3M}{r}-1\right]\bigg).
\end{split} \end{equation}
As before, we change the radial coordinate to $u=1/r$ and consider $2M=1$. Thus, the potential becomes
\begin{equation}\begin{split}
V_{\scriptscriptstyle{3/2}}&=\frac{u^2(1+\ell)(2+\ell)(3+\ell)\sqrt{1-u}}{2\big[u+(1+\ell)(3+\ell)\big]^2} \bigg(\frac{u^2}{2}\\ &+(1+\ell)(3+\ell)\left[(2+\ell)\sqrt{1-u}+\frac{3u}{2}-1\right]\bigg).
\end{split}
\end{equation}
Following the procedure implemented from Eq.~\eqref{Eq:TransN1} to Eq.~\eqref{Eq:AsymInfinite}, where we go from the Schr\"odinger-like equation to a differential equation suited to the numerical methods we are working with, the differential equation \eqref{eq:onda-s12} becomes
\begin{equation}\label{eq:onda-s32a}
\begin{split}
&R(\chi)\phi_{\scriptscriptstyle{3/2}}(\chi) + Q(\chi)\phi_{\scriptscriptstyle{3/2}}'(\chi) + P(\chi) \,\phi_{\scriptscriptstyle{3/2}}''(\chi)=0,
\end{split}
\end{equation}
where the coefficients $R(\chi)$, $Q(\chi)$, and $P(\chi)$ are given by
\begin{eqnarray} 
R(\chi)=
\,&& 2\big(1-\chi^2\big)\Big[6+11\ell+6\ell^2+\ell^3 +2\chi^5  \nonumber\\
&&+4\chi^4(2+\ell) +2\chi^3\left(3+4\ell+\ell^2\right) \nonumber\\
&&\!\!-\chi^2\left(2-7\ell-6\ell^2-\ell^3\right)  + 2\chi(2+\ell)^2\left(2+4\ell+\ell^2\right)\Big] \nonumber \\ 
&&\! -16\chi(2+\chi+\ell)^2\lambda\Big[i + 4 \left(2-\chi^2\right)\left(1-\chi^2\right)\lambda\Big],\\
Q(\chi) = && -\left(1-\chi^2\right)\left(2+\chi+\ell\right)^2\nonumber\\
&& \times \left[\left(1-\chi^2\right)^2-8\,i\,(1-4\chi^2+2\chi^4)\lambda\right],\\
 P(\chi)  =&& 
-\chi\left(1-\chi^2\right)^3\left(2+\chi+\ell\right)^2.
 \end{eqnarray}
where we have used the new coordinate $\chi^2=1-u$ to avoid square roots. Again, we get a quadratic eigenvalue problem, and the function $\phi_{\scriptscriptstyle{3/2}}(\chi)$ is regular in the interval $\chi\in [0,1]$. In Sec.~\ref{sec:numspin3/2} we solve Eq.~\eqref{eq:onda-s32a} by using the pseudo-spectral and AIM methods and compare our results against the results of Refs.~\cite{Shu:2005fw, Chen:2016qii}.

\subsection{Spin $5/2$ perturbations}

It is believed that the investigation of higher spin fields may shed some light on the understanding of fundamental physics, like on new unifying theories for the fundamental interactions, or on new phenomenology beyond the standard model.
The main motivation for investigating the spin $5/2$ field perturbation is the Rarita-Schwinger theory. Inspired by such a theory, the authors of Ref.~\cite{Shklyar:2009cx} computed some physical observable for the spin $\frac{5}{2}$-field. In this section we use the generic equation obtained in Ref.~\cite{Shu:2005fw}, specifically Eq.~(37), to determine the quasinormal frequencies of this perturbation field on the Schwarzschild black hole. The differential equation for the perturbations becomes so huge and for that reason we write it in Appendix \ref{Sec:Spin5f2Field}. The resulting equation is solved numerically by using the pseudo-spectral and AIM methods. The numerical results are displayed in Sec.~\ref{sec:numspin5/2}.

\section{The pseudo-spectral method}
\label{Sec:PseudoSpectral}

It is well known that Fourier method is appropriate to solve periodic problems, nevertheless it cannot be applied for nonperiodic problems due to the Gibbs phenomenon arising at the boundaries \cite{Arfken:379118}. An alternative method to solve nonperiodic problems is the pseudo-spectral method, which recently has being applied to solve differential equations numerically in many problems. The fact that the coordinate domain is not periodic, $u\in [0,1]$, allows us to use this method in our problem. Thus, the quadratic eigenvalue problem can be written in the form (using the notation of Ref.~\cite{Jansen:2017oag}). 
\begin{equation}\label{Eq:QuadraticEigenProb}
c_0(u,\lambda,\lambda^2)\phi(u)+c_1(u,\lambda,\lambda^2)\phi'(u)+c_2(u,\lambda,\lambda^2)\phi''(u)=0.
\end{equation}
The coefficients of this equation can be written as $c_j(u,\lambda,\lambda^2)=c_{j,0}(u)+\lambda\,c_{j,1}(u)+\lambda^2\,c_{j,2}(u)$, where $c_{j,0}(u)$, $c_{j,1}(u)$, and $c_{j,2}(u)$ are polynomials of $u$ only. 

The idea behind the pseudo-spectral method is to rewrite the regular function $\phi(u)$ in a base composed by cardinal functions $C_j(u)$, in the form
\begin{equation}\label{TrunSeries}
\phi_s(u)=\sum_{j=0}^{N}g(u_j)\,C_j(u),
\end{equation}
where $g(u)$ is a function of $u$. The next step is to evaluate the differential equation (including these functions) on a grid or collocation points. The best choice is the Gauss-Lobato grid given by
\begin{equation}\label{Eq:GaussLobato}
u_i=\frac{1}{2}\left(1\pm\cos{\left[\frac{i}{N}\pi\right]}\right),\quad i=0,1,2,\cdots, N
\end{equation}
Note that \eqref{Eq:GaussLobato} maps the interval $[-1,1]$ into $[0,1]$.

Evaluating on the grid, the polynomials of \eqref{Eq:QuadraticEigenProb} become elements of a matrix $c_j(u_i,\lambda,\lambda^2)=c_{j,0}(u_i)+\lambda\,c_{j,1}(u_i)+\lambda^2\,c_{j,2}(u_i)$. Then, the matrix representation of the quadratic eigenvalue problem \eqref{Eq:QuadraticEigenProb} can be written as
\begin{equation}\label{Eq:QuadraticEingenMatrix}
\begin{split}
\left(\tilde{M}_0+\tilde{M}_1\lambda+\tilde{M}_2\lambda^2\right)g=0,
\end{split}
\end{equation}
where 
\begin{equation}
\begin{split}
(\tilde{M}_{0})_{ji}=c_{0,0}(u_i)D_{ji}+c_{1,0}(u_i)D^{(1)}_{ji}+c_{2,0}(u_i)D^{(2)}_{ji},\\
(\tilde{M}_{1})_{ji}= c_{0,1}(u_i)D_{ji}+ c_{1,1}(u_i)D^{(1)}_{ji}+ c_{2,1}(u_i)D^{(2)}_{ji},\\
(\tilde{M}_{2})_{ji}= c_{0,2}(u_i)D_{ji}+ c_{1,2}(u_i)D^{(1)}_{ji}+ c_{2,2}(u_i)D^{(2)}_{ji},
\end{split}
\end{equation}
here $D_{ji}$, $D^{(1)}_{ji}$, and $D^{(2)}_{ji}$ represent the cardinal function and its derivatives. Defining $\tilde{g}=\lambda g$, the last equation may be written in the form
\begin{equation}
\begin{split}
\tilde{M}_0\,g+\left(\tilde{M}_1+\tilde{M}_2\lambda\right)\tilde{g}=0,&
\end{split}
\end{equation}
This is the first step to linearize the quadratic eigenvalue problem. For a generalization of this procedure see for instance Ref.~\cite{Tisseur:2017}. Therefore, the matrix representation of the eigenvalue problem may be written as
\begin{equation}\label{Eq:LinearEigenvalue2}
\left(M_0+M_1\,\lambda\right)\cdot\vec{g}=\mathbb{0},
\end{equation}
where we have defined the new matrices
\begin{equation}
M_0=
\begin{pmatrix}
\tilde{M}_0 & \tilde{M}_1 \\
\mathbb{0} & \mathbb{1}
\end{pmatrix},\,
M_1=
\begin{pmatrix}
\mathbb{0} & \tilde{M}_2 \\
-\mathbb{1} & \mathbb{0}
\end{pmatrix},\,
\vec{g}=
\begin{pmatrix}
g \\
\tilde{g}
\end{pmatrix}.
\end{equation}
Notice that ${M}_0$ and ${M}_1$ are $(N+1)\times (N+1)$ matrices and $\vec{g}$ is a $(N+1)-$dimensional vector with components $g_j=g(u_j)$, $j=0,\,1,...,N$. Finally, the QN frequencies are determined solving the linear eigenvalue problem \eqref{Eq:LinearEigenvalue2}. The last procedure was explained for a quadratic eigenvalue problem. This procedure can be easily extended for arbitrary order of the eigenvalue problem whenever the power of the frequency is an integer. However, if the value of the potential changes at the infinite spatial, as in the case of massive scalar field, see for instance \cite{Konoplya:2004wg}, the implementation of the pseudo-spectral method is not obvious because the power of the frequency turns out semi-integer such that it is not possible to write the eigenvalue problem in the form of Eq.~\eqref{Eq:QuadraticEingenMatrix}.

Having described how to calculate the eigenvalues, we need to specify the cardinal functions. We realized that these functions may depend on one or more Chebyshev polynomials of the first kind $T_{k}(u)$. In the following, we consider two forms for the cardinal functions. The first model considers one Chebyshev polynomial in the form
\begin{equation}
C_j(u)=T_{j}(u).
\end{equation}
We call this particular choice as pseudo-spectral I. The second model considers two Chebyshev polynomials \cite{Jansen:2017oag}
\begin{equation}
C_j(u)=\frac{2}{Np_j}\sum_{m=0}^{N}\frac{1}{p_m}T_m(u_j)T_m(u), \quad
\begin{cases}
p_0=2,\\
p_N=2,\\
p_j=1.
\end{cases}
\end{equation}
We call this choice pseudo-spectral II. It is worth mentioning that the pseudo-spectral method inevitably leads to the 
emergence of spurious solutions that do not have any physical 
meaning. To eliminate the spurious solutions we use the 
fact that the relevant QN frequencies do not depend on the number of
Chebyshev polynomials being considered. An additional check of consistency is to plot the function $\phi_s(u)$, which must satisfy the boundary conditions, i.e., regular at the horizon and divergent at the spatial infinity. Note that the problem of calculating QN frequencies does not depend on any initial guess, as the shooting method, for example. We get directly the frequencies using, for instance, Mathematica's built-in function Eigenvalues, or Eigensystem. One may consider this fact as an advantage in relation to other methods available in the literature.

The Chebyshev polynomials of the first kind $T_{k}(x)$ are defined in the interval $x\in [-1,1]$, and have special properties \cite{Arfken:379118}. Note also that the collocation points \eqref{Eq:GaussLobato} map this interval into the interval $[0,1]$.  In turn, the error associated with the pseudo-spectral method is of the order $\mathcal{O}\left(1/N^N\right)$ for sufficiently smooth regular functions \cite{Grandclement:2007sb}. For further discussions see for instance Ref.~\cite{Jansen:2017oag}.

\section{The Asymptotic Iteration Method}
\label{Sec:AIM}

The Asymptotic Iteration Method (AIM) is a numerical method recently proposed by Ciftci et. al.~\cite{Ciftci:2003} for solving homogeneous second order ordinary differential equations of the form
\begin{equation}
    y^{\prime\prime}(x) - \lambda_0(x)y^\prime(x) - s_0(x)y(x) = 0,
    \label{eq:aim_general_ode}
\end{equation}
where primes denote derivatives with respect to to the variable $x$ (that is defined over some interval that is not necessarily bounded), $\lambda_0(x) \neq 0$ and $s_0(x)$ are $C_\infty$. These equations can be found in many different areas of physics, such as the time independent Schr\"odinger equation in Quantum Mechanics,  or in General Relativity, such as the differential equations for black hole perturbations Eq.~\eqref{Eq:IntegerSpin} (where we can restore the first derivative if the standard radial coordinate is used instead of tortoise coordinates). Here we present a brief review concerning the AIM and discuss some implementation details of such a method. The AIM is based upon the following theorem:

\begin{theorem}
Let $\lambda_0$ and $s_0$ be functions of the variable $x \in (a,b)$ that are $C_\infty$ on the same interval. The differential equation~\eqref{eq:aim_general_ode} has a general solution of the form
\begin{multline}
    y(x) = \exp\left( -\int\alpha\ud t \right) \times\\
    \left[ C_2 + C_1 \int^{x} \exp \left( \int^{t} ( \lambda_0(\tau) + 2\alpha(\tau) )\ud \tau \right) \ud t \right]
    \label{eq:aim_general_solution}
\end{multline}
if for some $n>0$ the condition
\begin{equation}
    \alpha \equiv \frac{s_n}{\lambda_n} = \frac{s_{n-1}}{\lambda_{n-1}}
    \label{eq:aim_alpha_definition}
\end{equation}
or equivalently
\begin{equation}
    \delta \equiv s_n\lambda_{n-1} - \lambda_{n}s_{n-1} = 0
    \label{eq:aim_delta_definition}
\end{equation}
is satisfied, where
\begin{align}
    \lambda_k(x) \equiv & \lambda^\prime_{k-1}(x) + s_{k-1}(x) + \lambda_0(x)\lambda_{k-1}(x) \label{eq:aim_lambda_k}\\
    s_k(x) \equiv & s^\prime_{k-1}(x) + s_0\lambda_{k-1}(x) \label{eq:aim_sk}
\end{align}
with $k$ being a integer that ranges from $1$ to $n$.
\label{theo:aim_theorem}
\end{theorem}

From now on, we shall refer to the condition expressed by Eq.~\eqref{eq:aim_delta_definition} as the \emph{AIM quantization condition}. Provided that Theo.~\ref{theo:aim_theorem} is satisfied we can find both the eigenvalues and eigenvectors of the second order ODE using, respectively, Eq.~\eqref{eq:aim_delta_definition} and Eq.~\eqref{eq:aim_general_solution}. More specifically, the quasinormal modes of a perturbed black hole will be the complex frequency values $\omega$ that satisfy Eq.~\eqref{eq:aim_delta_definition} for any value of $x$.

Despite being quite general, the method presents a computational difficulty hidden in Eq.\eqref{eq:aim_lambda_k} and Eq.~\eqref{eq:aim_sk}: Not only the definition of the $n$-th coefficients are coupled and recursive, they also involve the derivatives of previous entries. In practice this means that to compute the quantization condition, Eq.~\eqref{eq:aim_delta_definition}, using $n$ iterations we end up computing the $n$-th derivatives of $\lambda_0$ and $s_0$ multiple times. Depending on the size of the original functions, the size and complexity of each coefficient can quickly spiral out of control. To address these issues, Cho et. al. have proposed in Ref.~\cite{Cho:2011sf} to instead of computing these coefficients directly, use a Taylor expansion of both $\lambda$ and $s$ around a point $\xi$ where the AIM is to be perform (we remind the reader that the results are independent of the choice of $\xi$), that is,
\begin{align}
    \lambda_n(\xi) = & \sum_{i=0}^{\infty}c^{i}_n(x - \xi)^i, \label{eq:taylor_lambda0}\\
    s_n(\xi) = & \sum_{i=0}^{\infty}d^{i}_n(x - \xi)^i, \label{eq:taylor_s0}
\end{align}
where $c^i_n$ and $d^i_n$ are the Taylor coefficients of the expansions of $\lambda_n$ and $s_n$ around $\xi$, respectively. By plugging Eqs.~\eqref{eq:taylor_lambda0} and \eqref{eq:taylor_s0} into Eqs.~\eqref{eq:aim_lambda_k} and Eq.~\eqref{eq:aim_sk} one gets

\begin{align}
    c^i_n = & (i+1)c^{i+1}_{n-1} + d^i_{n-1} + \sum_{k=0}^{i}c^k_0c^{i-k}_{n-1}, \label{eq:cin_def} \\
    d^i_n = & (i+1)d^{i+1}_{n-1} + \sum_{k=0}^{i}d^k_0c^{i-k}_{n-1}. \label{eq:din_def}
\end{align}
Finally, using Eqs.~\eqref{eq:cin_def} and \eqref{eq:din_def} the quantization condition, Eq.~\eqref{eq:aim_delta_definition}, becomes
\begin{equation}
    \delta \equiv d^0_n c^0_{n-1} - d^0_{n-1}c^0_n = 0.
    \label{eq:improved_delta}
\end{equation}

In order to better visualize and understand the improved algorithm, 
it is useful to arrange the $c^i_n$ (or $d^i_n$) coefficients as elements $c_{i,n}$ (or $d_{i,n}$) of a matrix $C$ (or $D$), where the index $i$ indicates the matrix row and the index $n$ represents the matrix column, i.e.,
\begin{equation*}
    C = 
    \begin{pmatrix}
        c_{0,0} & c_{0,1} & \cdots & c_{0,n-1} & c_{0,n} \\
        c_{1,0} & c_{1,1} & \cdots & c_{1,n-1} & c_{1,n} \\
        \vdots  & \vdots  & \ddots & \vdots    & \vdots  \\
        c_{i,0} & c_{i,1} & \cdots & c_{i,n-1} & c_{i,n} 
        \end{pmatrix}
    .
    \label{eq:matrix_model_of_coeffs}
\end{equation*}
Notice that, when using Eq.~\eqref{eq:improved_delta}, only the last and the before last top elements of the matrix (from left to right) are actually required, i.e., only the elements $c_{0,n-1}$ and $c_{0,n}$ are necessary. Note also that, by using Eqs.~\eqref{eq:cin_def} and \eqref{eq:din_def},  to compute an element in row $i$ and column $n$, one needs to have previously computed the column $n-1$ up to at least row $i+1$. In practice this means that the matrix $C$ ``grows diagonally'' and in order to compute $n$ columns one needs at least $i=n$ rows.  This formulation motivates us to view Eqs.~\eqref{eq:cin_def} and \eqref{eq:din_def} not as recursion relations, but as recipes for iteration, i.e., given the first column of the matrix $C$, one can use Eqs.~\eqref{eq:cin_def} and \eqref{eq:din_def} rewritten as
\begin{align}
    c^i_{n+1} = & (i+1)c^{i+1}_n + d^i_n + \sum_{k=0}^{i}c^k_0c^{i-k}_n, \label{eq:cin_iterative} \\
    d^i_{n+1} = & (i+1)d^{i+1}_n + \sum_{k=0}^{i}d^k_0c^{i-k}_n, \label{eq:din_iterative}
\end{align}
to compute the next column of the matrix.

With this insight, we can now devise an algorithm that performs $n$ iterations of the AIM. Remember that if $n$ iterations are to be performed, one needs at least $i=n$ rows of coefficients and thus we shall truncate the Taylor expansions at $i=n$. The algorithm steps are the following:

\begin{enumerate}
    \item Construct two arrays of size $n$ where the $i$-th element is $c^i_0$ (or $d^i_0$) where $i$ ranges from zero to $n$. We shall call these \texttt{icda} (initial $c$ data array) and \texttt{idda} (initial $d$ data array).
    
    \item Construct two arrays of size $n$ to contain the current column of $c$ (or $d$) indexes. We shall call these \texttt{ccda} (current $c$ data array) and \texttt{cdda} (current $d$ data array)
    
    \item Construct two arrays of size $n$ to contain the previous column of $c$ (or $d$) indexes. We shall call these \texttt{pcda} (previous $c$ data array) and \texttt{pdda} (previous $d$ data array).
    
    \item Initialize \texttt{ccda} with data from \texttt{icda} and \texttt{cdda} with data from \texttt{idda}.
    
    \item Perform $n$ AIM steps using the evolution Eqs.~\eqref{eq:cin_iterative} and \eqref{eq:din_iterative}. That is, repeat the following $n$ times:
    
    \begin{enumerate}
        \item Copy the content from \texttt{ccda} into \texttt{pcda}
        \item Copy the content from \texttt{cdda} into \texttt{pdda}
        \item Rewrite each element of \texttt{ccda} and \texttt{cdda} using Eqs.~\eqref{eq:cin_iterative} and \eqref{eq:din_iterative}, respectively.
    \end{enumerate}
    
    \item After $n$ iterations, the current and previous data array contain the sought coefficients. Apply the quantization condition, Eq.~\eqref{eq:improved_delta}, using the first indexes of each array (as they represent the $i=0$ coefficients). Explicitly, perform \texttt{cdda[1]*pcda[1] - pdda[1]*ccda[1]}\footnote{Given that Julia uses 1 based array indexes, we are also using 1 based arrays for the algorithmic description. This means \texttt{cdda[1]} refers to the first element of the \texttt{cdda} array and so on so forth.}.
    
    \item Finding the roots of the resulting expression from the last step yields the eigenvalues of the ODE (in the context of this work, the quasinormal modes).
\end{enumerate}

This implementation is realized in the Julia~\cite{BEZANSON:2017} package called \texttt{QuasinormalModes.jl}~\cite{SANCHES:2021}. The implementation makes use of an additional buffer array for each coefficient family in order to allow for thread based parallelization to take place during the main AIM loop. All AIM numerical results from this work were obtained with the aforementioned package.


\section{Numerical results for the QN frequencies}
\label{Sec:NumericalResults}

\subsection{Spin 0 QN frequencies}
\label{sec:numspin0}

Our numerical results for the quasinormal frequencies of the spin $0$ perturbation field are displayed in Table \ref{Tab:Spin0}. The first two column shows the data from the pseudo-spectral method with different numbers of interpolating polynomials, the third column shows the results form the AIM method, while the fourth and fifth columns are reproduction of the results from Refs.~\cite{Shu:2005fw, Konoplya:2004ip}, respectively. Notice that in Refs.~\cite{Shu:2005fw, Konoplya:2004ip} the WKB method was used to calculate the quasinormal frequencies. As it can be seen, the pseudo-spectral method I (calculated with $60$ polynomials) and pseudo-spectral method II (calculated with $40$ polynomials) provide results which are practically the same as those provided by the AIM, within six decimal places of precision. The numerical methods employed here provide more accurate results than those obtained by using the WKB approximation, and allows us to calculate additional frequencies for spin 0 fields not reported previously in the literature.

\begin{table*}[ht]
\centering
\begin{tabular}{l |c|c|c|c|c|c}
\hline 
$l$ & $n$ & \tlt{Pseudo-spectral}{I (60 Polynomials)} & \tlt{Pseudo-spectral}{II (40 polynomials)} & \tlt{AIM}{100 Iterations}  & Ref.~\cite{Shu:2005fw} &  Ref.~\cite{Konoplya:2004ip} \\ \hline\hline
0   & $0$ & $\pm 0.110455 -0.104896 i$                & $\pm 0.110455 -0.104896 i$                 & $\pm 0.110455 -0.104896 i$ & $0.1046-0.1152 i$ & $\pm 0.1105-0.1008i$    \\ \hline
1   & $0$ & $\pm 0.292936 -0.097660 i$                & $\pm 0.292936 -0.097660 i$                 & $\pm 0.292936 -0.097660 i$ & $0.2911-0.0980 i$ & $\pm 0.2929-0.0978i$    \\
    & $1$ & $\pm 0.264449 -0.306257 i$                & $\pm 0.264449 -0.306257 i$                 & $\pm 0.264449 -0.306257 i$ & ---               & $\pm 0.2645-0.3065i$    \\ \hline
2   & $0$ & $\pm 0.483644 -0.096759 i$                & $\pm 0.483644 -0.096759 i$                 & $\pm 0.483644 -0.096759 i$ & $0.4832-0.0968 i$ & $\pm 0.4836-0.0968i$    \\
    & $1$ & $\pm 0.463851 -0.295604 i$                & $\pm 0.463851 -0.295604 i$                 & $\pm 0.463851 -0.295604 i$ & $0.4632-0.2958 i$ & $\pm 0.4638-0.2956i$    \\
    & $2$ & $\pm 0.430544 -0.508558 i$                & $\pm 0.430544 -0.508558 i$                 & $\pm 0.430544 -0.508558 i$ & ---               & $\pm 0.4304-0.5087i$    \\ \hline
3   & $0$ & $\pm 0.675366 -0.096500 i$                & $\pm 0.675366 -0.096500 i$                 & $\pm 0.675366 -0.096500 i$ & $0.6752-0.0965 i$ & ---                     \\
    & $1$ & $\pm 0.660671 -0.292285 i$                & $\pm 0.660671 -0.292285 i$                 & $\pm 0.660671 -0.292285 i$ & $0.6604-0.2923 i$ & ---                     \\
    & $2$ & $\pm 0.633626 -0.496008 i$                & $\pm 0.633626 -0.496008 i$                 & $\pm 0.633626 -0.496008 i$ & $0.6348-0.4941 i$ & ---                     \\
    & $3$ & $\pm 0.598773 -0.711221 i$                & $\pm 0.598773 -0.711221 i$                 & $\pm 0.598773 -0.711221 i$ & ---               & ---                     \\ \hline
4   & $0$ & $\pm 0.867416 -0.096392 i$                & $\pm 0.867416 -0.096392 i$                 & $\pm 0.867416 -0.096392 i$ & $0.8673-0.0964 i$ & ---                     \\
    & $1$ & $\pm 0.855808 -0.290876 i$                & $\pm 0.855808 -0.290876 i$                 & $\pm 0.855808 -0.290876 i$ & $0.8557-0.2909 i$ & ---                     \\
    & $2$ & $\pm 0.833692 -0.490325 i$                & $\pm 0.833692 -0.490325 i$                 & $\pm 0.833692 -0.490325 i$ & $0.8345-0.4895 i$ & ---                     \\
    & $3$ & $\pm 0.803288 -0.697482 i$                & $\pm 0.803288 -0.697482 i$                 & $\pm 0.803288 -0.697482 i$ & $0.8064-0.6926 i$ & ---                     \\
    & $4$ & $\pm 0.767733 -0.914019 i$                & $\pm 0.767733 -0.914019 i$                 & $\pm 0.767733 -0.914019 i$ & ---               & ---                     \\
\hline\hline
\end{tabular}
\caption{
Quasinormal frequencies of the spin $0$ perturbations compared against the results of Refs.~\cite{Shu:2005fw, Konoplya:2004ip}.
}
\label{Tab:Spin0}
\end{table*}

\subsection{Spin 1  QN frequencies}
\label{sec:numspin1}

Our numerical results for the quasinormal frequencies for spin $1$ fields are displayed in Table \ref{Tab:Spin1} compared against the results of Refs.~\cite{Shu:2005fw, Konoplya:2004ip}, where the WKB method was used. The first two columns show the data from the pseudo-spectral method with different numbers of interpolating polynomials, the third column shows the results form the AIM method, while the fourth and fifth columns are reproduction of the results from Refs.~\cite{Shu:2005fw, Konoplya:2004ip}, respectively. As it can be seen from the table, the pseudo-spectral method I (calculated with $60$ polynomials) and pseudo-spectral method II (calculated with $40$ polynomials) provide results which are practically identical to those provided by the AIM. The numerical methods we are working with provide more accurate results than those obtained by using the WKB approximation, and allows us to present additional frequencies for spin 1 fields not reported previously in the literature.

\begin{table*}[ht]
\centering
\begin{tabular}{l |c|c|c|c|c|c}
\hline 
$l$ & $n$ & \tlt{Pseudo-spectral}{I (60 Polynomials)} & \tlt{Pseudo-spectral}{II (40 polynomials)} & \tlt{AIM}{100 Iterations}     & Ref.~\cite{Shu:2005fw} &  Ref.~\cite{Konoplya:2004ip} \\ \hline\hline
1   & $0$ & $\pm 0.248263-0.092488 i$                 & $\pm 0.248263 -0.092488 i$                 & $\pm 0.248263 -0.092488 i$    & $0.2459-0.0931i$ & $\pm 0.2482-0.0926i$    \\
    & $1$ & $\pm 0.214515-0.293668 i$                 & $\pm 0.214515 -0.293667 i$                 & $\pm 0.214515 -0.293668 i$    & ---              & $\pm 0.2143-0.2941i$    \\ \hline
2   & $0$ & $\pm 0.457596-0.095004 i$                 & $\pm 0.457595 -0.095004 i$                 & $\pm 0.457596 -0.095004 i$    & $0.4571-0.0951i$ & $\pm 0.4576-0.0950i$    \\
    & $1$ & $\pm 0.436542-0.290710 i$                 & $\pm 0.436542 -0.290710 i$                 & $\pm 0.436542 -0.290710 i$    & $0.4358-0.2910i$ & $\pm 0.4365-0.2907i$    \\
    & $2$ & $\pm 0.401187-0.501587 i$                 & $\pm 0.401187 -0.501587 i$                 & $\pm 0.401187 -0.501587 i$    & ---              & $\pm 0.4009-0.5017i$    \\ \hline
3   & $0$ & $\pm 0.656899-0.095616 i$                 & $\pm 0.656899 -0.095616 i$                 & $\pm 0.656899 -0.095616 i$    & $0.6567-0.0956i$ & $\pm 0.6569-0.0956i$    \\
    & $1$ & $\pm 0.641737-0.289728 i$                 & $\pm 0.641737 -0.289728 i$                 & $\pm 0.641737 -0.289728 i$    & $0.6415-0.2898i$ & $\pm 0.6417-0.2897i$    \\
    & $2$ & $\pm 0.613832-0.492066 i$                 & $\pm 0.613832 -0.492066 i$                 & $\pm 0.613832 -0.492066 i$    & $0.6151-0.4901i$ & $\pm 0.6138-0.4921i$    \\
    & $3$ & $\pm 0.577919-0.706331 i$                 & $\pm 0.577919 -0.706331 i$                 & $\pm 0.577919 -0.706330 i$    & ---              & $\pm 0.5775-0.7065i$    \\ \hline
4   & $0$ & $\pm 0.853095-0.095860 i$                 & $\pm 0.853095 -0.095860 i$                 & $\pm 0.853095 -0.095810 i$    & $0.8530-0.0959i$ & ---                     \\
    & $1$ & $\pm 0.841267-0.289315 i$                 & $\pm 0.841267 -0.289315 i$                 & $\pm 0.841267 -0.289315 i$    & $0.8411-0.2893i$ & ---                     \\
    & $2$ & $\pm 0.818728-0.487838 i$                 & $\pm 0.818728 -0.487838 i$                 & $\pm 0.818728 -0.487838 i$    & $0.8196-0.4870i$ & ---                     \\
    & $3$ & $\pm 0.787748-0.694242 i$                 & $\pm 0.787748 -0.694242 i$                 & $\pm 0.787748 -0.694242 i$    & $0.7909-0.6892i$ & ---                     \\
    & $4$ & $\pm 0.751549-0.910242 i$                 & $\pm 0.751549 -0.910242 i$                 & $\pm 0.751549 -0.910242 i$    & ---              & ---                     \\
\hline\hline
\end{tabular}
\caption{
Quasinormal frequencies of the spin $1$ perturbations  compared against the results of Refs.~\cite{Shu:2005fw, Konoplya:2004ip}.
}
\label{Tab:Spin1}
\end{table*}

\subsection{Spin 2  QN frequencies}
\label{sec:numspin2}

Our numerical results for the quasinormal frequencies for spin $2$ fields are displayed in Table \ref{Tab:Spin2} compared against the results of Refs.~\cite{Shu:2005fw, Konoplya:2004ip}, where the WKB method was employed. The first two columns show the data from the pseudo-spectral method with different numbers of interpolating polynomials, the third column shows the results form the AIM method, while the fourth and fifth columns are reproduction of the results from Refs.~\cite{Shu:2005fw, Konoplya:2004ip}, respectively. As it can be seen from the table, the pseudo-spectral method I and pseudo-spectral method II provide results which are practically equal to those provided by the AIM. The numerical results we are working with provide more accurate results than those obtained by using the WKB approximation. It is worth mentioning that, in this case, we obtain additional solutions to the eigenvalue problem which do not represent gravitational waves, see the discussion in Appendix \ref{Sec:AddDiscu} for more details on this point.

\begin{table*}[!htbp]
\centering
\begin{tabular}{l |c|c|c|c|c|c}
\hline 
$l$ & $n$ & \tlt{Pseudo-spectral}{I (60 Polynomials)} & \tlt{Pseudo-spectral}{II (40 polynomials)} & \tlt{AIM}{100 Iterations}  & Ref.~\cite{Shu:2005fw} &  Ref.~\cite{Konoplya:2004ip} \\ \hline\hline
2   & $0$ & $\pm 0.373672-0.088962i$                  & $\pm 0.373672 -0.088962 i$                 & $\pm 0.373672 -0.088962 i$ & $0.3730-0.0891i$       & $\pm 0.3736-0.0890i$    \\
    & $1$ & $\pm 0.346711-0.273915i$                  & $\pm 0.346711 -0.273915 i$                 & $\pm 0.346711 -0.273915 i$ & $0.3452-0.2746i$       & $\pm 0.3463-0.2735i$    \\
    & $2$ & $\pm 0.301053-0.478277i$                  & $\pm 0.301053 -0.478277 i$                 & $\pm 0.301053 -0.478277 i$ & ---                    & $\pm 0.2985-0.4776i$    \\ \hline
3   & $0$ & $\pm 0.599443-0.092703i$                  & $\pm 0.599443 -0.092703 i$                 & $\pm 0.599443 -0.092703 i$ & $0.5993-0.0927i$       & $\pm 0.5994-0.0927i$    \\
    & $1$ & $\pm 0.582644-0.281298i$                  & $\pm 0.582644 -0.281298 i$                 & $\pm 0.582644 -0.281298 i$ & $0.5824-0.2814i$       & $\pm 0.5826-0.2813i$    \\
    & $2$ & $\pm 0.551685-0.479093i$                  & $\pm 0.551685 -0.479093 i$                 & $\pm 0.551685 -0.479027 i$ & $0.5532-0.4767i$       & $\pm 0.5516-0.4790i$    \\
    & $3$ & $\pm 0.511962-0.690337i$                  & $\pm 0.511962 -0.690337 i$                 & $\pm 0.511962 -0.690337 i$ & ---                    & $\pm 0.5111-0.6905i$    \\ \hline
4   & $0$ & $\pm 0.809178-0.094164i$                  & $\pm 0.809178 -0.094164 i$                 & $\pm 0.809178 -0.094164 i$ & $0.8091-0.0942i$       & $\pm 0.8092-0.0942i$    \\
    & $1$ & $\pm 0.796632-0.284334i$                  & $\pm 0.796632 -0.284334 i$                 & $\pm 0.796632 -0.284334 i$ & $0.7965-0.2844i$       & $\pm 0.7966-0.2843i$    \\
    & $2$ & $\pm 0.772710-0.479908i$                  & $\pm 0.772710 -0.479908 i$                 & $\pm 0.772710 -0.479908 i$ & $0.7736-0.4790i$       & $\pm 0.7727-0.4799i$    \\
    & $3$ & $\pm 0.739837-0.683924i$                  & $\pm 0.739837 -0.683924 i$                 & $\pm 0.739837 -0.683924 i$ & $0.7433-0.6783i$       & $\pm 0.7397-0.6839i$    \\
    & $4$ & $\pm 0.701516-0.898239i$                  & $\pm 0.701516 -0.898239 i$                 & $\pm 0.701516 -0.898239 i$ & ---                    & $\pm 0.7006-0.8985i$    \\
\hline\hline
\end{tabular}
\caption{
Quasinormal frequencies of spin $2$ perturbations compared against the results of Refs.~\cite{Konoplya:2004ip, Shu:2005fw}.
}
\label{Tab:Spin2}
\end{table*}

\subsection{Spin $1/2$ QN frequencies}
\label{sec:numspin1/2}

Our numerical results for the quasinormal frequencies for spin $1/2$ fields are displayed in Table \ref{Tab:Spin1/2} compared against results available in the literature. The first two columns show the data from the pseudo-spectral method with different numbers of interpolating polynomials, the third column shows the results form the AIM method, while the fourth and fifth columns are reproduction of the results from Refs.~\cite{Cho:2003qe, Shu:2005fw}, respectively.
As it can be seen, the results obtained using the pseudo-spectral I and II are in perfect agreement with the results obtained using the AIM within the decimal places considered. Note that the numerical methods we are working with provide more accurate results than the results reported in Refs.~\cite{Cho:2003qe, Shu:2005fw}, where the authors employed the WKB approximation. Note that we also show additional frequencies not reported previously in the literature, for example for $\ell=1$ and $n=1$. 

It is worth pointing out that we also found purely imaginary frequencies, that arise when investigating the quasinormal modes in the limit of large $\ell$. Our numerical results are displayed in Table \ref{Tab:PurelyImSpin3/2}, where we show the first five purely imaginary frequencies.
As it is seen from the table, the agreement between the numerical methods is perfect for low overtones but it gets worse 
for higher overtones.  It is worth mentioning that these results are in perfect agreement with the analytic solution, $M\omega=-i{n}/{4},\quad n\to \infty$, obtained in Refs.~\cite{Cho:2005yc, Khriplovich:2005wf}, see also references therein.

\begin{table*}[!htbp]
\centering
\begin{tabular}{l |c|c|c|c|c|c}
\hline 
$l$ & $n$ & \tlt{Pseudo-spectral}{I (60 Polynomials)} & \tlt{Pseudo-spectral}{II (40 polynomials)} & \tlt{AIM}{100 Iterations}  & Ref.~\cite{Shu:2005fw} &  Ref.~\cite{Cho:2003qe} \\ \hline\hline
0   & $0$ & $\pm 0.182963 -0.096982 i$                & $\pm 0.182963 -0.096982 i$                 & $\pm 0.182963 -0.096824 i$ & ---                    & ---                     \\ \hline
1   & $0$ & $\pm 0.380037 -0.096405 i$                & $\pm 0.380037 -0.096405 i$                 & $\pm 0.380037 -0.096405 i$ & $0.3786-0.0965 i$      & $0.379 -0.097i$         \\
    & $1$ & $\pm 0.355833 -0.297497 i$                & $\pm 0.355833 -0.297497 i$                 & $\pm 0.355833 -0.297497 i$ & ---                    & ---                     \\ \hline
2   & $0$ & $\pm 0.574094 -0.096305 i$                & $\pm 0.574094 -0.096305 i$                 & $\pm 0.574094 -0.096305 i$ & $0.5737-0.0963 i$      & $0.574 -0.096i$         \\
    & $1$ & $\pm 0.557015 -0.292715 i$                & $\pm 0.557015 -0.292715 i$                 & $\pm 0.557015 -0.292715 i$ & $0.5562-0.2930 i$      & $0.556 -0.293i$         \\
    & $2$ & $\pm 0.526607 -0.499695 i$                & $\pm 0.526607 -0.499695 i$                 & $\pm 0.526607 -0.499695 i$ & ---                    & ---                     \\ \hline
3   & $0$ & $\pm 0.767355 -0.096270 i$                & $\pm 0.767355 -0.096270 i$                 & $\pm 0.767355 -0.096270 i$ & $0.7672-0.0963 i$      & $0.767 -0.096i$         \\
    & $1$ & $\pm 0.754300 -0.290968 i$                & $\pm 0.754300 -0.290968 i$                 & $\pm 0.754300 -0.290968 i$ & $0.7540-0.2910 i$      & $0.754 -0.291i$         \\
    & $2$ & $\pm 0.729770 -0.491910 i$                & $\pm 0.729770 -0.491910 i$                 & $\pm 0.729770 -0.491910 i$ & $0.7304-0.4909 i$      & $0.730 -0.491i$         \\
    & $3$ & $\pm 0.696913 -0.702293 i$                & $\pm 0.696913 -0.702293 i$                 & $\pm 0.696913 -0.702293 i$ & ---                    & ---                     \\ \hline
4   & $0$ & $\pm 0.960293 -0.096254 i$                & $\pm 0.960293 -0.096254 i$                 & $\pm 0.960293 -0.096254 i$ & $0.9602-0.0963 i$      & $0.960 -0.096i$         \\
    & $1$ & $\pm 0.949759 -0.290148 i$                & $\pm 0.949759 -0.290148 i$                 & $\pm 0.949759 -0.290148 i$ & $0.9496-0.2902 i$      & $0.950 -0.290i$         \\
    & $2$ & $\pm 0.929494 -0.488116 i$                & $\pm 0.929494 -0.488116 i$                 & $\pm 0.929494 -0.488116 i$ & $0.9300-0.4876 i$      & $0.930 -0.488i$         \\
    & $3$ & $\pm 0.901129 -0.692520 i$                & $\pm 0.901129 -0.692520 i$                 & $\pm 0.901129 -0.692520 i$ & $0.9036-0.6892 i$      & $0.904 -0.689i$         \\
    & $4$ & $\pm 0.867043 -0.905047 i$                & $\pm 0.867008 -0.905066 i$                 & $\pm 0.867043 -0.905047 i$ & ---                    & ---                     \\
\hline\hline
\end{tabular}
\caption{
Quasinormal frequencies of the spin $1/2$ perturbations compared against the results of Refs.~\cite{Cho:2003qe, Shu:2005fw}.
}
\label{Tab:Spin1/2}
\end{table*}

\begin{table}[t]
\centering
\begin{tabular}{c|c|c}
\hline 
\tlt{Pseudo-spectral}{I (60 Polynomials)} & \tlt{Pseudo-spectral}{II (40 polynomials)} & \tlt{AIM}{100 Iterations}  \\ \hline\hline
$-0.250000i$                & $-0.250000i$                 & $-0.250000i$  \\ \hline
$-0.500000i$                & $-0.500000i$                 & $-0.500000i$  \\ \hline
$-0.750000i$                & $-0.750000i$                 & $-0.750000i$  \\ \hline
$-1.000000i$                & $-1.000000i$                 & $-1.000031i$  \\ \hline
$-1.2499998i$               & $-1.250000i$                 & $-1.246550i$  \\
\hline\hline
\end{tabular}
\caption{
Purely imaginary frequencies for spin $1/2$ perturbations. The numerical values of such frequencies are exactly the same as for the purely imaginary frequencies arising in the QNM of spin 3/2 perturbations.
}
\label{Tab:PurelyImSpin3/2}
\end{table}

\subsection{Spin $3/2$ QN frequencies}
\label{sec:numspin3/2}

Our numerical results for the quasinormal frequencies for spin $3/2$ fields are displayed in Table \ref{Tab:Spin3/2} compared against results available in the literature.   The first two columns show the data from the pseudo-spectral method with different numbers of interpolating polynomials, the third column shows the results form the AIM method, while the fourth and fifth columns are reproduction of the results from Refs.~\cite{Shu:2005fw,Chen:2016qii}, respectively. As it can be seen, the results obtained by using the pseudo-spectral I and II are in perfect agreement with the results obtained using the AIM. We also realize that these results are in very good agreement with the results reported in Ref.~\cite{Chen:2016qii}, where the authors also employed the IAM, and with the results from Ref.~\cite{Shu:2005fw}, where the authors employed the WKB approximation.

As in the spin $1/2$ field perturbations, we also find purely imaginary frequencies for the spin $3/2$ field. The numerical results are displayed in Table \ref{Tab:PurelyImSpin3/2} for the three routines we are working with. Such frequencies arise when investigating the quasinormal modes in the limit of large $\ell$. Notice that these results are also in agreement with the analytic solutions obtained in Refs.~\cite{Cho:2005yc, Khriplovich:2005wf}. It is worth pointing out that the numerical values of these purely imaginary frequencies are exactly the same for spin $1/2$ and $3/2$ fields. We do not have an explanation for this fact, maybe it is just a coincidence. Notice also that these frequencies can be written as fractions, multiples of $1/4$, i.e., $1/4$, $2/4$, $3/4$, $4/4$, $\sim 5/4$, $\cdots$.

\begin{table*}[b]
\centering
\begin{tabular}{l |c|c|c|c|c|c}
\hline 
$l$ & $n$ & \tlt{Pseudo-spectral}{I (60 Polynomials)} & \tlt{Pseudo-spectral}{II (40 polynomials)} & \tlt{AIM}{100 Iterations}  & Ref.~\cite{Shu:2005fw} &  Ref.~\cite{Chen:2016qii} \\ \hline\hline
0   & $0$ & $\pm 0.311292 -0.090087 i$                & $\pm 0.311292 -0.090087 i$                 & $\pm 0.311292 -0.090087 i$ & ---                    & $0.3112 -0.0902 i$        \\ \hline
1   & $0$ & $\pm 0.530048 -0.093751 i$                & $\pm 0.530048 -0.093751 i$                 & $\pm 0.530048 -0.093751 i$ & ---                    & $0.5300 -0.0937 i$        \\
    & $1$ & $\pm 0.511392 -0.285423 i$                & $\pm 0.511392 -0.285423 i$                 & $\pm 0.511392 -0.285423 i$ & ---                    & $0.5113 -0.2854 i$        \\ \hline
2   & $0$ & $\pm 0.734750 -0.094878 i$                & $\pm 0.734750 -0.094878 i$                 & $\pm 0.734750 -0.094878 i$ & $\pm 0.7346 -0.0949 i$ & $0.7347 -0.0948 i$        \\
    & $1$ & $\pm 0.721047 -0.286906 i$                & $\pm 0.721047 -0.286906 i$                 & $\pm 0.721047 -0.286906 i$ & $\pm 0.7206 -0.2870 i$ & $0.7210 -0.2869 i$        \\
    & $2$ & $\pm 0.695287 -0.485524 i$                & $\pm 0.695287 -0.485524 i$                 & $\pm 0.695287 -0.485524 i$ & ---                    & $0.6952 -0.4855 i$        \\ \hline
3   & $0$ & $\pm 0.934364 -0.095376 i$                & $\pm 0.934364 -0.095376 i$                 & $\pm 0.934364 -0.095376 i$ & $\pm 0.9343 -0.0954 i$ & $0.9343 -0.0953 i$        \\
    & $1$ & $\pm 0.923502 -0.287560 i$                & $\pm 0.923502 -0.287560 i$                 & $\pm 0.923502 -0.287560 i$ & $\pm 0.9233 -0.2876 i$ & $0.9235 -0.2875 i$        \\
    & $2$ & $\pm 0.902599 -0.483957 i$                & $\pm 0.902599 -0.483957 i$                 & $\pm 0.902599 -0.483957 i$ & $\pm 0.9031 -0.4835 i$ & $0.9025 -0.4839 i$        \\
    & $3$ & $\pm 0.873342 -0.687024 i$                & $\pm 0.873343 -0.687024 i$                 & $\pm 0.873342 -0.687024 i$ & ---                    & $0.8732 -0.6870 i$        \\ \hline
4   & $0$ & $\pm 1.131530 -0.095640 i$                & $\pm 1.131530 -0.095640 i$                 & $\pm 1.131530 -0.095640 i$ & $\pm 1.1315 -0.0956 i$ & $1.1315 -0.0956 i$        \\
    & $1$ & $\pm 1.122523 -0.287908 i$                & $\pm 1.122523 -0.287908 i$                 & $\pm 1.122523 -0.287908 i$ & $\pm 1.1224 -0.2879 i$ & $1.1225 -0.2879 i$        \\
    & $2$ & $\pm 1.104976 -0.483096 i$                & $\pm 1.104976 -0.483096 i$                 & $\pm 1.104976 -0.483096 i$ & $\pm 1.1053 -0.4828 i$ & $1.1049 -0.4830 i$        \\
    & $3$ & $\pm 1.079852 -0.683000 i$                & $\pm 1.079852 -0.683000 i$                 & $\pm 1.079852 -0.683000 i$ & $\pm 1.0817 -0.6812 i$ & $1.0798 -0.6829 i$        \\
    & $4$ & $\pm 1.048599 -0.889113 i$                & $\pm 1.048596 -0.889115 i$                 & $\pm 1.048599 -0.889113 i$ & ---                    & $1.0484 -0.8890 i$        \\
\hline\hline
\end{tabular}
\caption{
Quasinormal frequencies of spin $3/2$ perturbations compared against the results of Refs.~\cite{Chen:2016qii, Shu:2005fw}.
}
\label{Tab:Spin3/2}
\end{table*}

\subsection{Spin $5/2$ QN frequencies}
\label{sec:numspin5/2}

Our numerical results for the quasinormal frequencies for spin $5/2$ fields are displayed in Table \ref{Tab:Spin5/2}. As it can be seen, the results obtained by using the pseudo-spectral I and II are in perfect agreement with the results obtained using the AIM within the decimal places considered. It is worth mentioning that we present  here the quasinormal frequencies for spin $5/2$ perturbation fields for the very first time

In turn, we also found purely imaginary frequencies. Our numerical results are displayed in Table \ref{Tab:PurelyImSpin5/2}. It turns out that these results satisfy the sequence $1/8$, $\sim 3/8$, $\sim 5/8$, $\sim 7/8$, $\cdots$, in general, 
\begin{equation}
M\omega=-i\frac{(2n+1)}{8},\qquad n=0,1,2,3,\cdots. 
\end{equation}
As in spin $1/2$ and $3/2$ perturbations, these frequencies arise when we were investigating the frequencies in the limit of large $\ell$. As it can be seen, the discrepancy of both methods increases as the imaginary frequency gets negative.

Finally, differently from the quasinormal frequencies with real and imaginary parts, the purely imaginary frequencies represent damping solutions, while the frequencies with real and imaginary parts represent oscillatory solutions being damped by the imaginary part of the frequency. This is easily understood because the solution goes like $\sim e^{i\omega\,t}=e^{i(\omega_{Re}-i\,\omega_{Im})\,t}=e^{\omega_{Im}\,t}\cos{\left(\omega_{Re}\,t\right)}$.

\begin{table*}[!htbp]
\centering
\begin{tabular}{l |c|c|c|c}
\hline 
$l$ & $n$ & \tlt{Pseudo-spectral}{I (60 Polynomials)} & \tlt{Pseudo-spectral}{II (40 polynomials)} & \tlt{AIM}{100 Iterations}  \\ \hline\hline
0   & $0$ & $\pm 0.462727-0.092578i$                & $\pm 0.462727-0.092578i$                 & $0.462727 - 0.092577 i$ \\ \hline
1   & $0$ & $\pm 0.687103-0.094566i$                & $\pm 0.687103-0.094566i$                 & $0.687103 - 0.094566 i$ \\
    & $1$ & $\pm 0.670542-0.285767i$                & $\pm 0.670542-0.285767i$                 & $0.670542 - 0.285767 i$ \\ \hline
2   & $0$ & $\pm 0.897345-0.095309i$                & $\pm 0.897345-0.095309i$                 & $0.897345 - 0.095309 i$ \\
    & $1$ & $\pm 0.884980-0.287266i$                & $\pm 0.884980-0.287266i$                 & $0.884980 - 0.287266 i$ \\
    & $2$ & $\pm 0.861109-0.483113i$                & $\pm 0.861109-0.483113i$                 & $0.861109 - 0.483113 i$ \\ \hline
3   & $0$ & $\pm 1.101190-0.095648i$                & $\pm 1.101190-0.095648i$                 & $1.101190 - 0.095648 i$  \\
    & $1$ & $\pm 1.091300-0.287886i$                & $\pm 1.091300-0.287886i$                 & $1.091300 - 0.287886 i$ \\
    & $2$ & $\pm 1.071999-0.482895i$                & $\pm 1.071999-0.482895i$                 & $1.071999 - 0.482895 i$ \\
    & $3$ & $\pm 1.044272-0.682307i$                & $\pm 1.044272-0.682307i$                 & $1.044272 - 0.682307 i$ \\ \hline
4   & $0$ & $\pm 1.301587-0.095829i$                & $\pm 1.301587-0.095829i$                 & $1.301587 - 0.095829 i$ \\
    & $1$ & $\pm 1.293328-0.288184i$                & $\pm 1.293328-0.288184i$                 & $1.293328 - 0.288184 i$ \\
    & $2$ & $\pm 1.277107-0.482604i$                & $\pm 1.277107-0.482604i$                 & $1.277107 - 0.482604 i$ \\
    & $3$ & $\pm 1.253526-0.680366i$                & $\pm 1.253526-0.680366i$                 & $1.253526 - 0.680366 i$ \\
    & $4$ & $\pm 1.223513-0.882554i$                & $\pm 1.223512-0.882553i$                 & $1.223513 - 0.882554 i$ \\
\hline\hline
\end{tabular}
\caption{
Quasinormal frequencies of spin $5/2$ perturbations.
}
\label{Tab:Spin5/2}
\end{table*}

\begin{table}[t]
\centering
\begin{tabular}{c|c|c}
\hline 
\tlt{Pseudo-spectral}{I (60 Polynomials)} & \tlt{Pseudo-spectral}{II (40 polynomials)} & \tlt{AIM}{100 Iterations}  \\ \hline\hline
$-0.125000i$                & $-0.125000i$                 & $-0.125000i$  \\ \hline
$-0.375602i$                & $-0.375602i$                 & $-0.378659i$  \\ \hline
$-0.626877i$                & $-0.626877i$                 & $-0.623931i$  \\ \hline
$-0.878946i$                & $-0.878948i$                 & $-0.907374i$  \\
\hline\hline
\end{tabular}
\caption{
Purely imaginary frequencies of spin $5/2$ perturbations.
}
\label{Tab:PurelyImSpin5/2}
\end{table}

\section{QNMs for $\ell \gg1$ and $n\gg 1$}
\label{Sec:QNMsLargel}

It is also interesting to calculate the quasinormal frequencies in the limit of large $\ell$, where analytic solutions are available in the literature to compare with. The analytic solutions we obtained in Ref.~\cite{Ferrari-1984}, such that the real and imaginary parts of the frequencies are given by
\begin{equation}\label{Eq:AnalyticFrequencies}
\begin{split}
M\omega_{\text{Re}}=\,&\frac{1}{3\sqrt{3}}\left(\ell+\frac{1}{2}\right),\\
M\omega_{\text{Im}}=\,&\frac{1}{3\sqrt{3}}\left(n+\frac{1}{2}\right).
\end{split}
\end{equation}
Before comparing our numerical results against the analytic solutions, it is worth mentioning that these analytic solutions were obtained for integer spin perturbations. We do not expected that these results can be applied for semi-integer field perturbations, in principle. 

\begin{figure}[ht!]
\centering
\includegraphics[width=7cm]{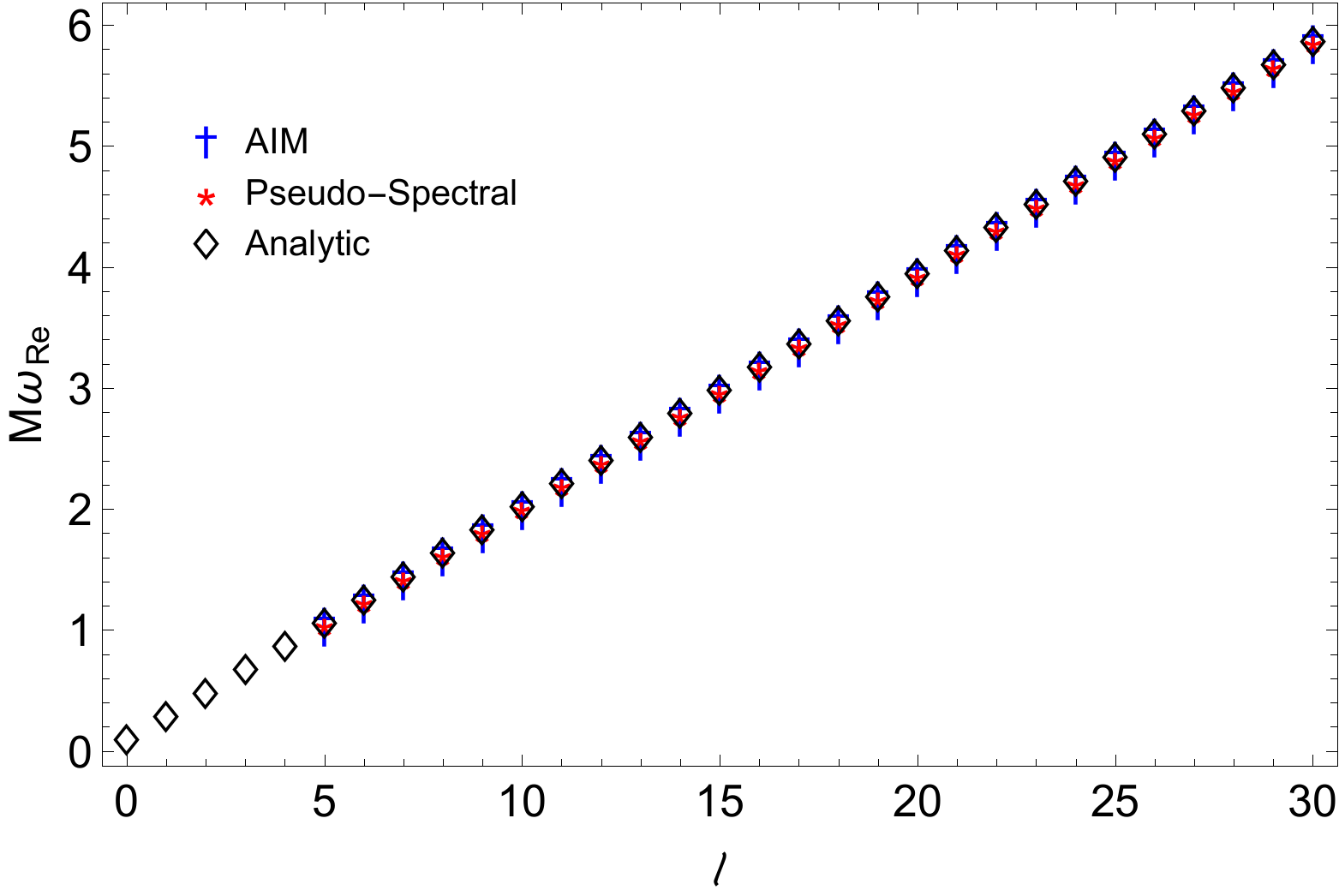}\\
\includegraphics[width=7cm]{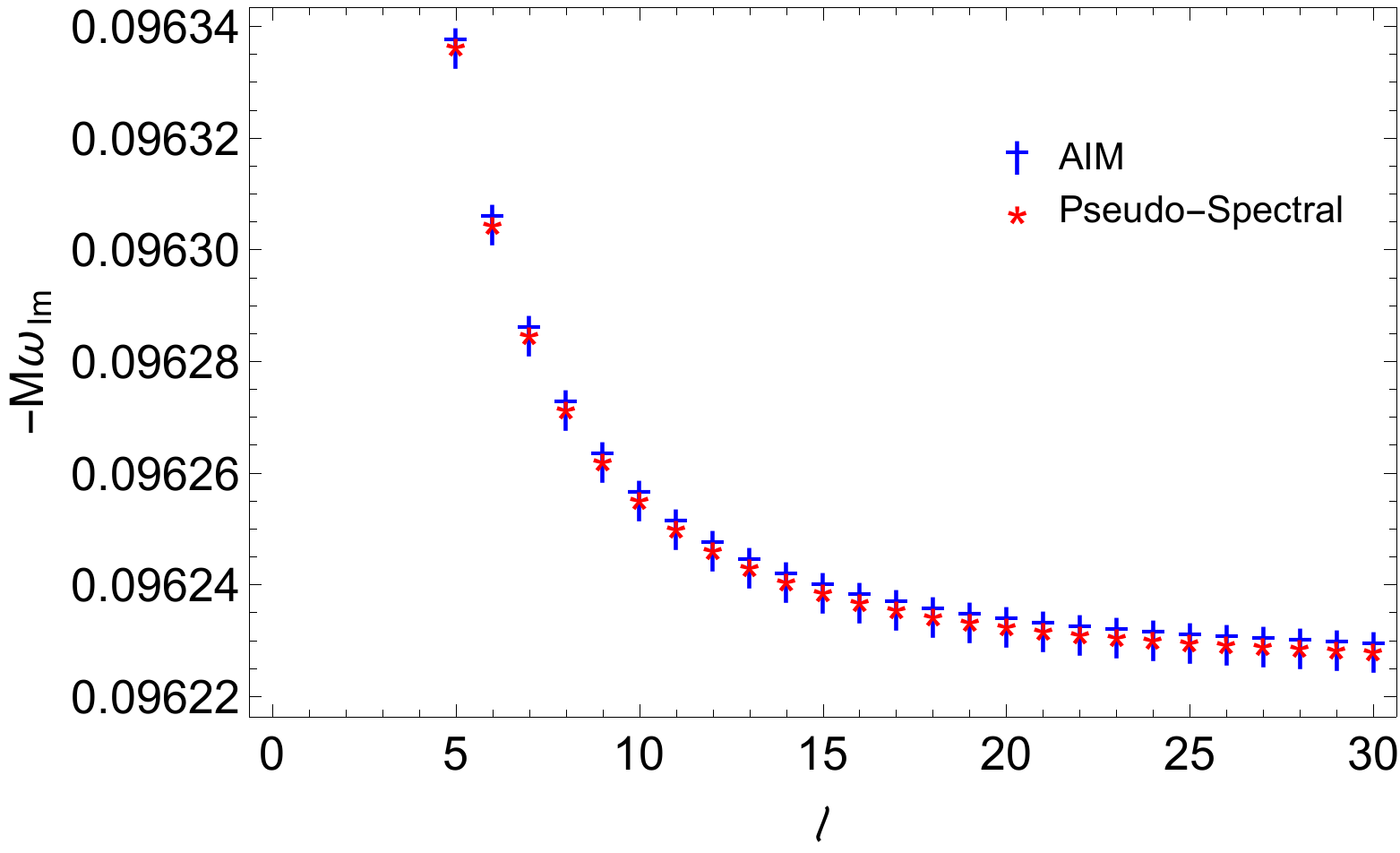}
\caption{The figure shows the real and imaginary parts of the frequency as a function of $\ell$ for $n=0$ considering $s=0$ field. Dagger represents the AIM results, while asterisk the pseudo-spectral and diamond analytic results of Eq.~\eqref{Eq:AnalyticFrequencies}.}
\label{Fig:LWS0}
\end{figure}

\begin{figure}[ht!]
\centering
\includegraphics[width=7cm]{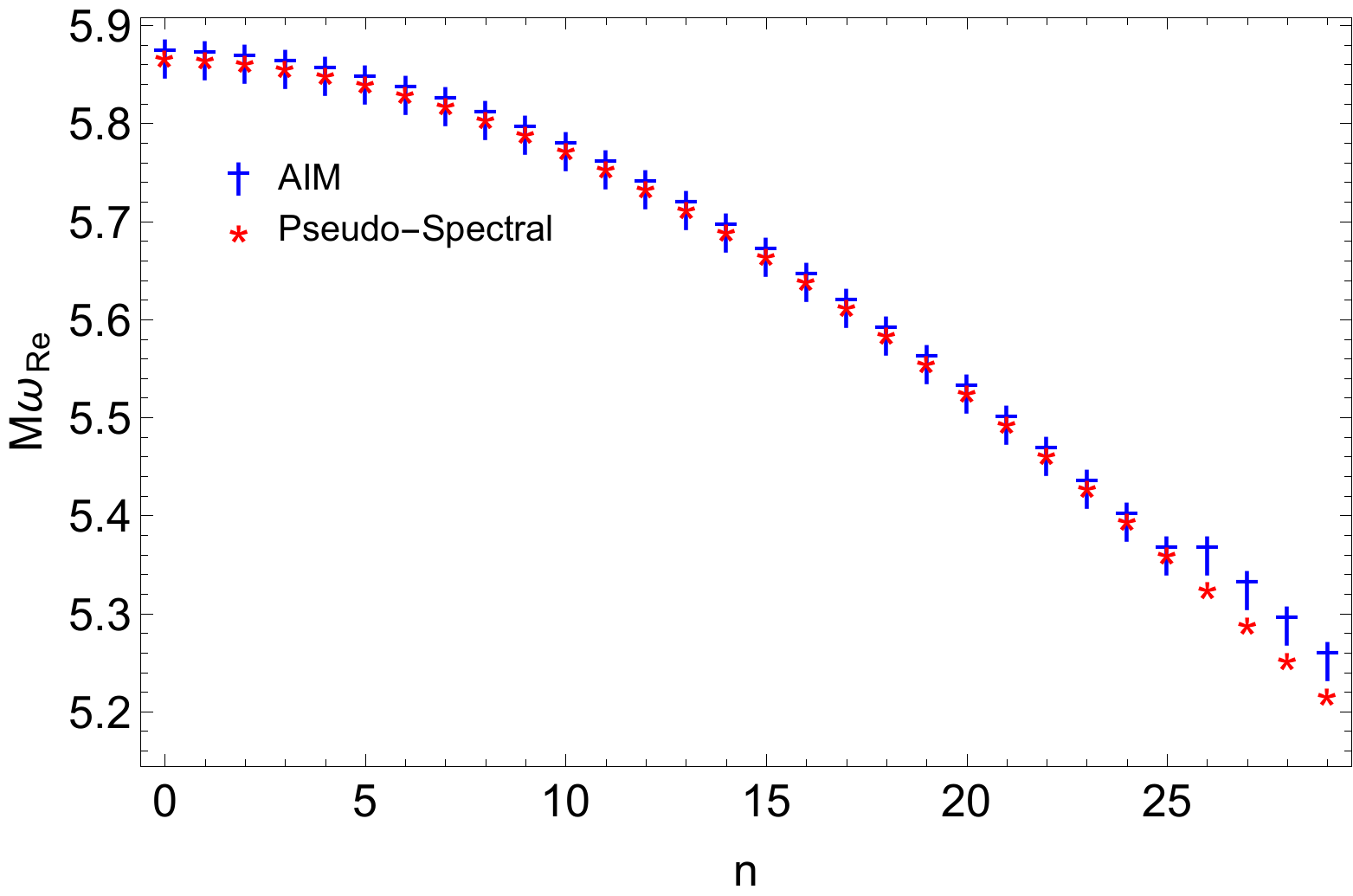}\\
\includegraphics[width=7cm]{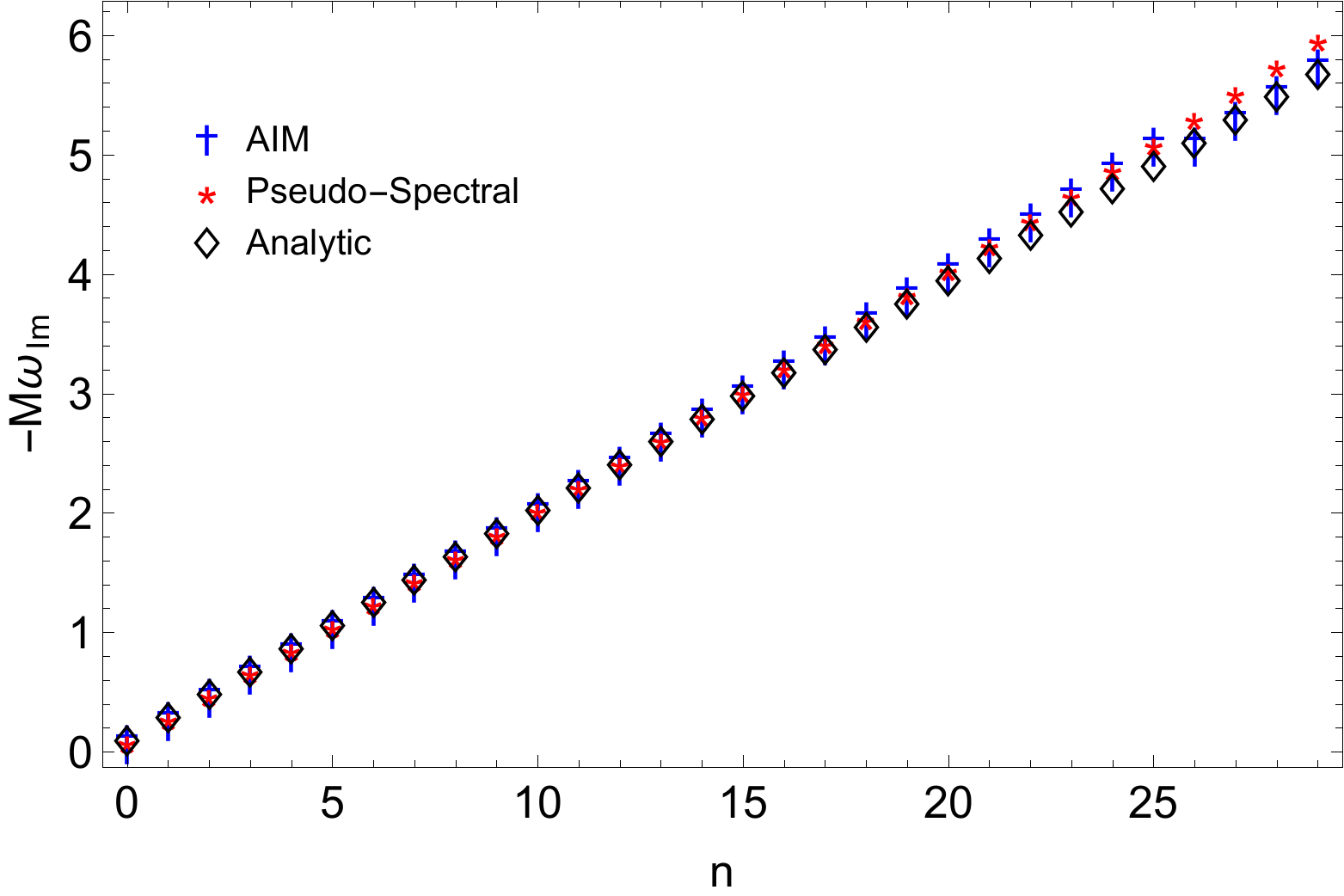}
\caption{The figure shows the real and imaginary parts of the frequency as a function of $n$ for $\ell=30$ considering $s=0$ field. Dagger represents the AIM results, while asterisk the pseudo-spectral and diamond analytic results of Eq.~\eqref{Eq:AnalyticFrequencies}.}
\label{Fig:nWS0}
\end{figure}

Let us now compare our numerical results against the analytic solutions \eqref{Eq:AnalyticFrequencies}. Our numerical results for $s=0$ field are displayed in Figs.~\ref{Fig:LWS0} and \ref{Fig:nWS0}, as a function of $\ell$ and $n$, respectively. As it can be seen, the real part as a function of $\ell$ fits very well the analytic result \eqref{Eq:AnalyticFrequencies}. Meanwhile, the imaginary part of the frequency as a function of $n$ also fits very well the analytic solution.

\begin{figure}[ht!]
\centering
\includegraphics[width=7cm]{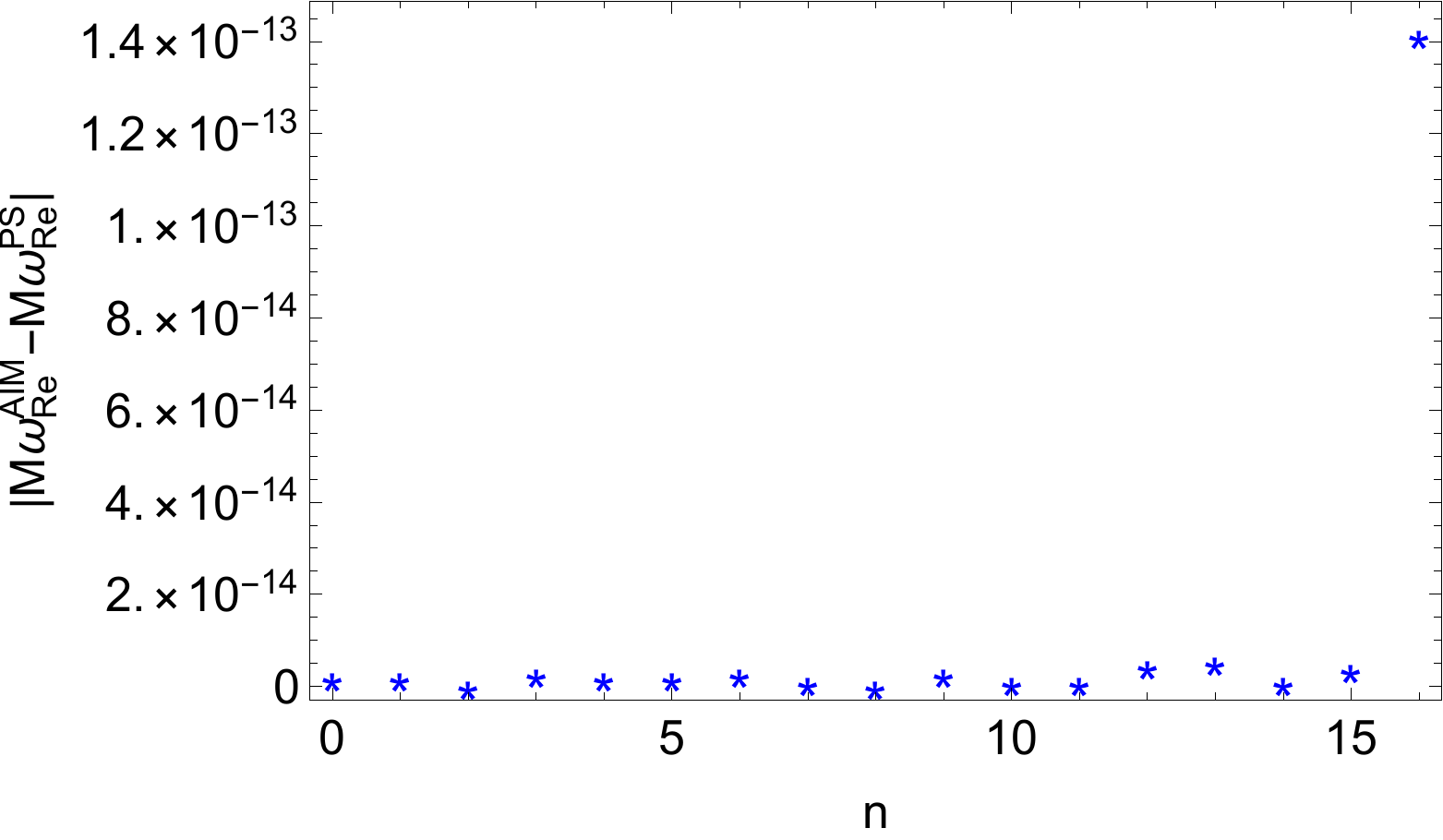}\\
\includegraphics[width=7cm]{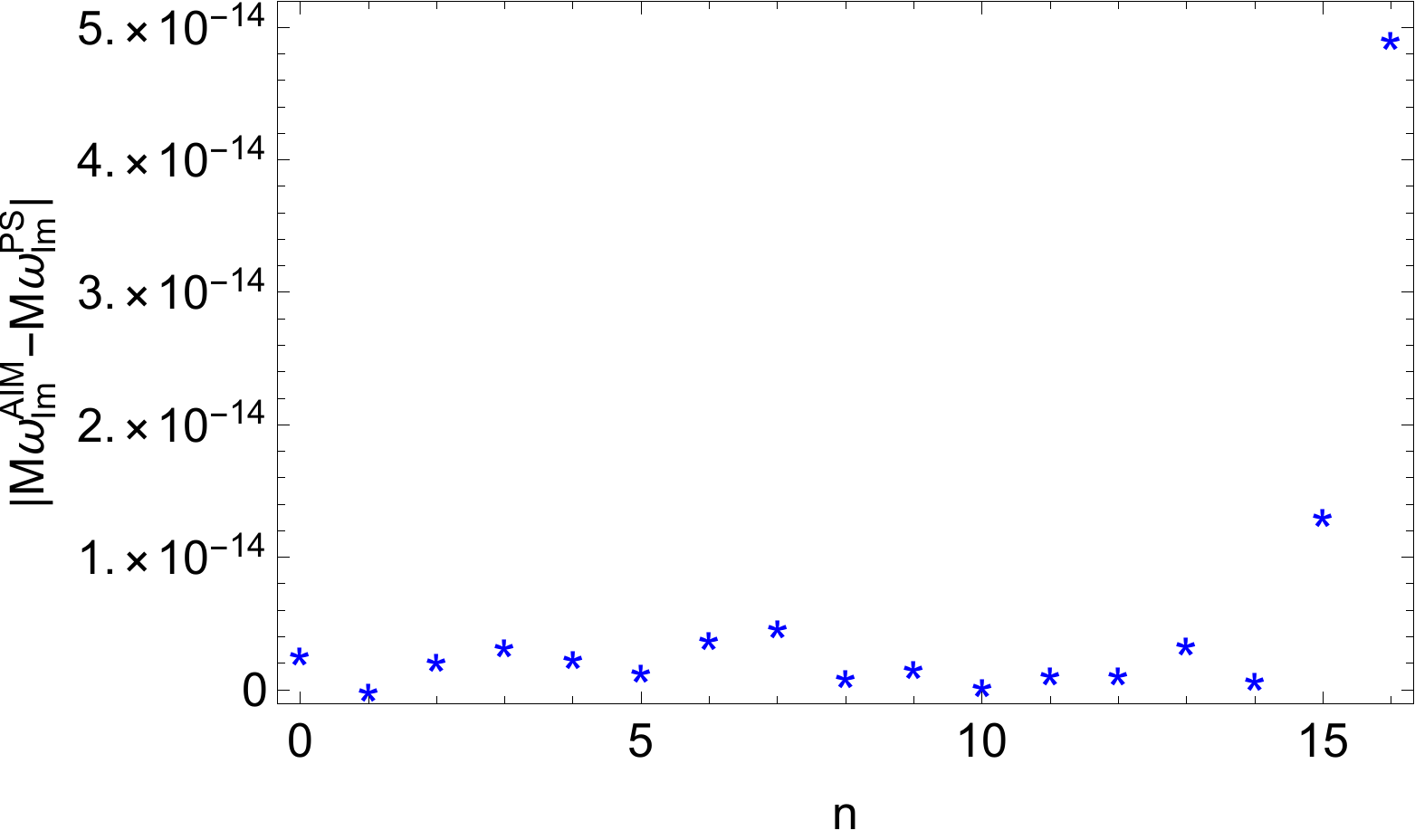}
\caption{The figure shows the difference of the real and imaginary parts of the frequency in logarithmic scale as a function of $n$ obtained in both, the AIM and Pseudo-Scalar methods for spin 0 perturbation.
}
\label{Fig:nLogWS0}
\end{figure}

It is worth pointing out that the numerical results obtained using the pseudo-spectral and AIM are in agreement as seen in these figures. To observe the numerical difference we plotted the difference of the frequencies $|M\omega^{\text{AIM}}_{\text{Re}}-M\omega^{\text{PS}}_{\text{Re}}|$ and $|M\omega^{\text{AIM}}_{\text{Im}}-M\omega^{\text{PS}}_{\text{Im}}|$ in logarithmic scale and displayed the results in Fig.~\ref{Fig:nLogWS0}. It is seen that the difference between the numerical results is very small and slightly increases with the increasing of $n$.

\begin{figure}[ht!]
\centering
\includegraphics[width=7cm]{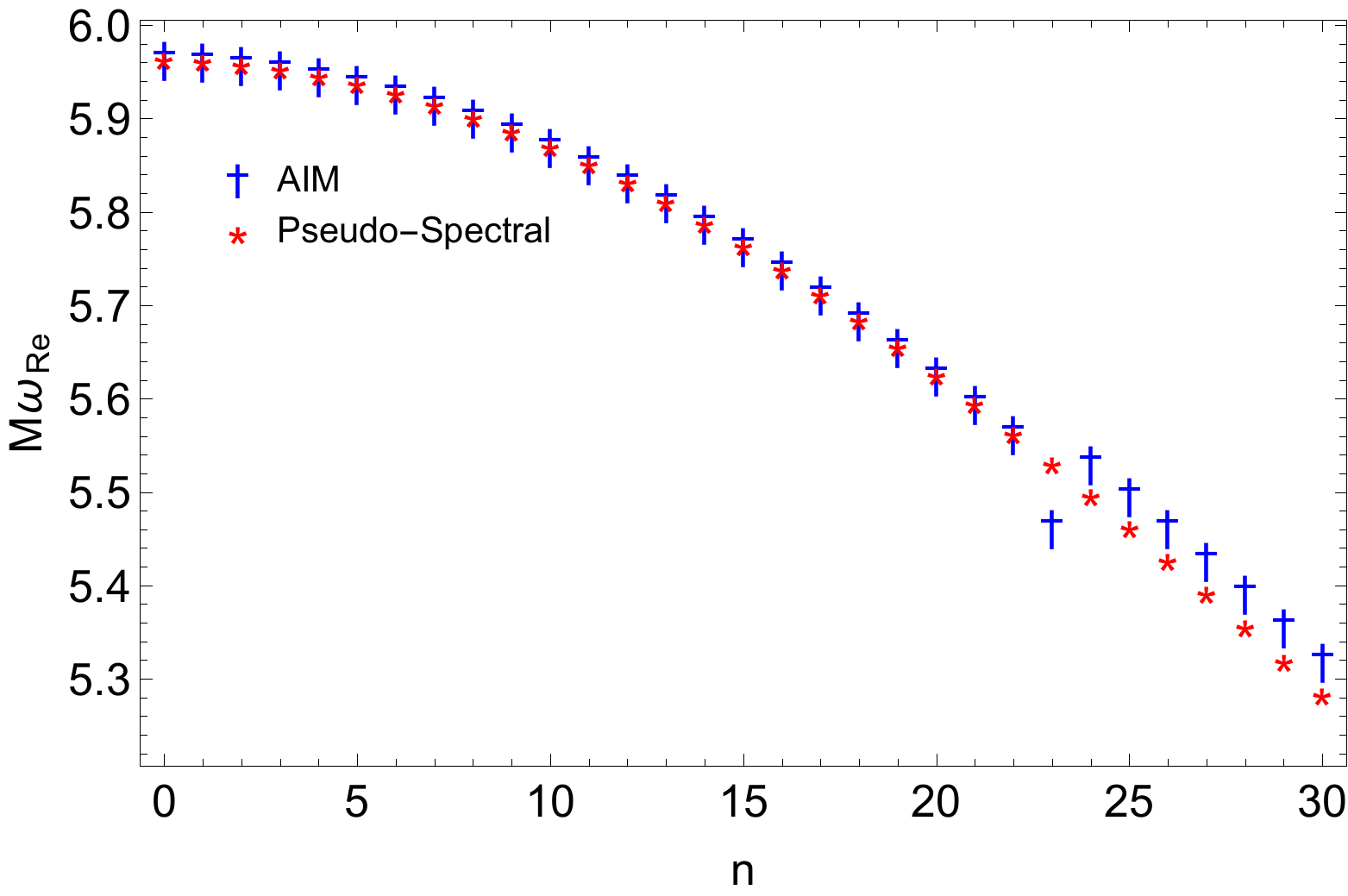}\\
\includegraphics[width=7cm]{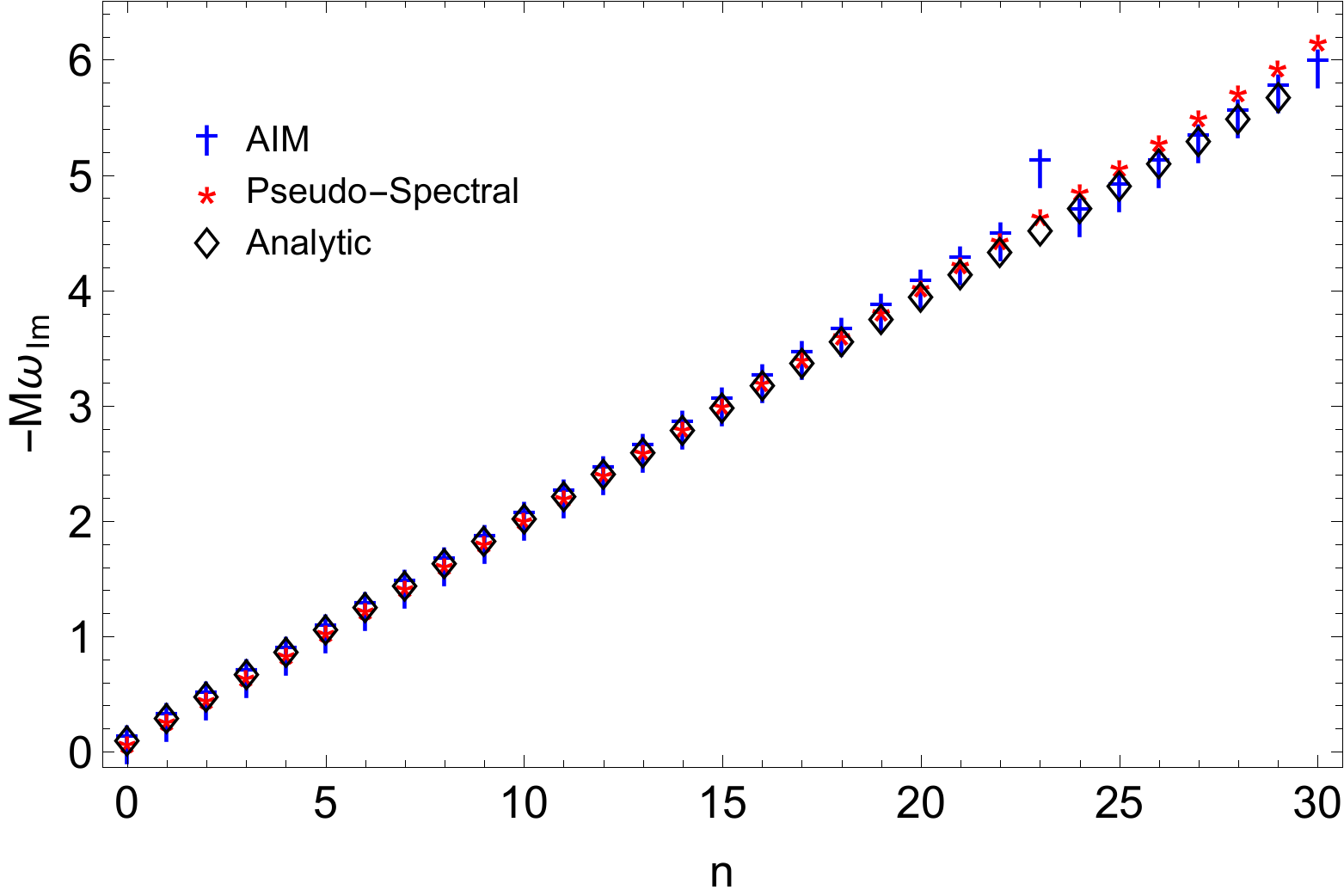}
\caption{The figure shows the real and imaginary parts of the frequency as a function of $n$ for $\ell=30$ considering $s=1/2$ field. Dagger represents the AIM results, while asterisk the pseudo-spectral and diamond analytic results of Eq.~\eqref{Eq:AnalyticFrequencies}.}
\label{Fig:nWS0p5}
\end{figure}

In turn, our numerical results for spin $1/2$ field are displayed in Fig~\ref{Fig:nWS0p5}. As it can be seen, the numerical and analytic results are in agreement. This means that the analytic results are also valid for perturbations for spin $1/2$.

\section{Discussion and Conclusions}
\label{Sec:Conclusions}

In this paper we calculated the quasinormal frequencies for spin $0$, $1/2$, $1$, $3/2$, $2$ and $5/2$ fields on the Schwarzschild black hole in asymptotically flat space-time. We have employed the pseudo-spectral and AIM methods. The main difference between these methods lies in the way they solve the eigenvalue problem. While the pseudo-spectral method expand the solution in a base of cardinal functions, the AIM calculates the roots of a characteristic polynomial. In doing so, the results obtained for a spin zero field applying the pseudo-spectral method are in good agreement with the results obtained by applying the AIM and also in good agreement with literature values.

We displayed results with six decimal places in Tables~\ref{Tab:Spin0} - \ref{Tab:Spin2}. The results show that both methods are in agreement at least up to the sixth decimal place. We point out that the quasinormal frequencies obtained for spin $0$, $1$, and $2$ fields do not bring any additional information to what we known from the literature. In turn, for spin $1/2$ and $3/2$ fields our results obtained for both methods are in agreement, these results are also in agreement with results available in the literature. For these fields, we have obtained additional frequencies which are purely imaginary, see Table~\ref{Tab:PurelyImSpin3/2}. We also calculated the quasinormal frequencies for a spin $5/2$ field, which are displayed in Table~\ref{Tab:Spin5/2}. We have observed that the pseudo-spectral method and the AIM show good agreement. Additional frequencies, which are purely imaginary, were also obtained and displayed in Table~\ref{Tab:PurelyImSpin5/2}. Hence, the pseudo-spectral method calculated a set of quasinormal frequencies, while the AIM used a guess to start looking for solutions. Our main conclusion is that both methods complemented each other in the task for calculating the QNM frequencies.

Additional comments concerning the numerical analysis may be of interest. The use of $60$ polynomials for the pseudo-spectral method I and $40$ polynomials for the pseudo-spectral method II are the minimum necessary to reach the precision one desires in the calculation, say 30 decimal places in such a case. An additional issue concerning the pseudo-spectral method is related to the reason why such a method does not yield, for instance, the 30 expected eigen-frequencies for $\ell=30$. We tried increasing the number of polynomials but, with fixed precision, were unable to go beyond 16 solutions. We realized that increasing the number of polynomials and the precision we are able to reach 30 eigen-frequencies. Similar difficulties are also present in the AIM.

In the next stage of the present work we are going to address the problem of calculating QNMs for massive test fields, extending the analysis to other black hole solutions like Reissnerd-Nordstr\"om, Kerr and Kerr-Newman black holes, as well as considering the contribution of the cosmological constant.

\acknowledgments

We would like acknowledge Alex S. Miranda for useful discussions along the development of this work. 
L.~T.~S. and A. D. D. Masa are partly founded by Coordena\c{c}\~ao de Aperfei\c{c}oamento de Pessoal de N\'{i}vel Superior (CAPES), Brazil, Finance Code 001. L.~A.~H.~M. is partially founded by the Universidade Estadual da Regi\~ao Tocantina do Maranh\~ao (UEMASUL, Brazil). V. T. Z. thanks financial support from Conselho  Nacional de Desenvolvimento Cient\'\i fico
e Tecnol\'ogico (CNPq), Brazil, Grant No.~309609/2018-6, and from CAPES, Brazil, Grant No. 88887.310351/2018-00.

\appendix

\section{Fundamental equations for the spin $5/2$ perturbations}
\label{Sec:Spin5f2Field}

In this appendix we write additional details of the spin $5/2$ field. First, we write the effective potential $V_{\scriptscriptstyle{5/2}}(u)$ that appears in the Schr\"odinger-like equation when written in terms of the variable $u=1/r$. It assumes the form
\begin{widetext}
\begin{equation}
\begin{split}
V_{\scriptscriptstyle{5/2}}&=\frac{u^2f}{4S(u)}\bigg\{4\bigg[75L_2\left(23+3L_3\right)u+9L_2^3u^6 +L_2^2\left[25L_4-9u^7\Big(9+L_3\{3+u^2-3u\}+u\left\{3+6u^2-13u\right\}\Big)\right]\\
&-450u^2\left[2u-\left(3+\ell\right)^2\right]\bigg]-15u^2f^{1/2}\bigg[2u\left(35\sqrt{7+L_3}-221+3L_3^2\left(\sqrt{7+L_3}-6u-16\right)+4L_3^3(u-1)\right.\\
&\left.+14u\left(5\sqrt{7+L_3}-98\right)+L_3\left(22\sqrt{7+L_3}-189-456u+14u\sqrt{7+L_3}\right)\right)\\
&+L_2\Big[12-20\sqrt{7+L_3}+9u^3(33+46u)+L_3\left(141u^3-4\sqrt{7+L_3}\right)\Big]\bigg]\bigg\},
\end{split}
\end{equation}
\end{widetext}
where, to simplify the expressions, we introduced the notation
\begin{align*}
S(u)=\,&\left[5\left(L_2+6u\right)+3L_2u^3f^{3/2}(u)\right]^2,\\
L_1=\,&11+\ell(6+\ell),\\
L_2=\,&(1+\ell)(5+\ell),\\
L_3=\,&\ell(6+\ell),\\
L_4=\,&(2+\ell)(4+\ell).
\end{align*}
Following the procedure explained in Section \ref{Sec:Spin1f2}, we transform the Schr\"odinger-like equation into a equation suitable to apply the pseudo-spectral method. Moreover, 
in order to avoid square roots we have used the variable $\chi^2=1-u$, such that the final differential equation is
\begin{equation} \label{eq:spin5/2}
R(\chi)\phi_{\scriptscriptstyle{5/2}}(\chi)+Q(\chi)\phi_{\scriptscriptstyle{5/2}}'(\chi)+
P(\chi)\phi_{\scriptscriptstyle{5/2}}''(\chi)=0,
\end{equation}
where $P(\chi)$, $Q(\chi)$, and $R(\chi)$ are given by
\begin{widetext}
\begin{equation}
\begin{split}
R(\chi)=&\chi\left(1-\chi^2\right)\bigg[100L_2L_4-800L_2L_4\left(\chi^2-1\right)+60\Big[10(17+L_3)+L_2\chi\left(\sqrt{7+L_3}(5+L_3)-3\right)\Big]\left(\chi^2-1\right)^2\\
&+15L_2\chi\left(305+133L_3\right)\left(\chi^2-1\right)^5+36L_2^2\left(10+3L_3\right)\left(\chi^2-1\right)^7+108L_2^2L_3\left(\chi^2-1\right)^8\\
&+36L_2^2(L_3-10)(\chi^2-1)^9-180L_2^2(\chi^2-1)^{10}+18L_2(2(5+L_3)^2-305\chi)(\chi^2-1)^6\\
&-60\chi\left[35\sqrt{7+L_3}-636+L_3\left(7\sqrt{7+L_3}-208+L_3\left(2L_3-7\right)\right)\right]\left(\chi^2-1\right)^4  \bigg]\\
&-30\chi\left(1-\chi^2\right)^4\Big(240+\chi\left[35\sqrt{7+L_3}-221-L_3\left(189-22\sqrt{7+L_3}+L_3\Big\{48-3\sqrt{7+L_3}+4L_3\Big\}\right)\right]\Big)\\
&-16\lambda\chi\Big[i+4\lambda\left(2-3\chi^2+\chi^4\right)\Big]\,t(\chi),\\
Q(\chi)=&\,\left(\chi^2-1\right)\Big[1+\chi^2\left(\chi^2-2\right)\left(1-16i\lambda\right)-8i\lambda\Big]\,t(\chi)\\
P(\chi)=&\,\chi\big(\chi^2-1\big)^3\,t(\chi),
\end{split}
\end{equation}

\end{widetext}
where $t(\chi)$ is an auxiliary coefficient given by
$t(\chi)=\left[30\left(\chi^2-1\right)+L_2\left(3\chi^9-9\chi^7+9\chi^5-3\chi^3-5\right)\right]^2$.
\section{Further comments on the numerical QN frequencies}
\label{Sec:AddDiscu}

In this appendix we comment on additional details found when analysing the quasinormal frequencies of the Schwarzschild black hole by means pseudo the AIM and pseudo-spectral methods for integer spin perturbation fields $s=2$ and $s=1$.

For completeness, we started our search for solutions to the eigenvalue problem for $s=2$ by setting $\ell=0$ in Eq.~\eqref{Eq:IntegerSpin3}, we do not get any solution. Then, we set $\ell=1$ and we get the solutions $M\omega_0=\pm 0.110455-0.104896i$ and $M\omega_1=\pm 0.086158-0.348079i$, corresponding to $n=0$ and $n=1$, respectively. It is also interesting to point out that the solution for $\ell=1$ and $n=0$ arises in both methods employed in this work, while the solution for $\ell=1$ and $n=1$ arises in the asymptotic iteration method alone. As a consistency check, we solved the same problem by employing the Leaver continued fraction method \cite{leaver1985analytic} and obtained the same results as from the AIM. However, as investigated by Regge-Wheeler \cite{Zerilli-1970}, and Zerilli \cite{Zerilli-1970} modes with $\ell=1$ represent an addition of angular momentum to the metric given by Eq.~\eqref{EqMetric}. Hence, such modes do not generate gravitational waves.

On the other hand, investigating the electromagnetic perturbation, $s=1$, for $\ell=0$ we obtained
\begin{equation}
M\omega_n=i\frac{(n+1)}{4},\qquad n=0,1,2,\cdots
\end{equation}
in both methods.

\bibliographystyle{apsrev4-2}
\bibliography{ref}

\end{document}